\documentclass[apj]{emulateapj}
\usepackage{psfig}

\shorttitle{Structure around CNELGs}
\shortauthors{Barton et al.}

\begin{document}
\title{A Search for Low Surface Brightness Structure Around Compact Narrow 
Emission Line Galaxies\altaffilmark{1}}
\author{\sc Elizabeth J. Barton\altaffilmark{2},
Liese van Zee\altaffilmark{3}, and
Matthew A. Bershady\altaffilmark{4}}

\altaffiltext{1}{Based on observations obtained at the
Canada-France-Hawaii Telescope which is operated by the National
Research Council of Canada, the Institut National des Sciences
de l'Univers of the Centre National de la 
Recherche Scientifique of France, and the University of Hawaii.}

\altaffiltext{2}{Center for Cosmology, Department of Physics and
Astronomy, University of California, Irvine,  CA 92697-4575
(email: ebarton@uci.edu)}

\altaffiltext{3}{Indiana University, Department of Astronomy, 727
E. 3rd Street, Bloomington, IN 47405 (email:vanzee@astro.indiana.edu)}

\altaffiltext{4}{Department of Astronomy, University of Wisconsin, 475
North Charter Street, Madison, WI 53706 (email: mab@astro.wisc.edu)}

\begin{abstract}

As the most extreme members of the rapidly evolving faint blue galaxy
population at intermediate redshift, the compact narrow emission line
galaxies (CNELGs) are intrinsically luminous ($-22 \lesssim {\rm M_B}
\lesssim -18$) with narrow emission linewidths ($30 \lesssim \sigma
\lesssim$ 125 km s$^{-1}$). Their nature is heavily debated: they may
be low-mass starbursting galaxies that will fade to present-day dwarf
galaxies or bursts of star formation temporarily dominating the flux
of more massive galaxies, possibly related to {\it in situ} bulge
formation or the formation of cores of galaxies.  We present deep,
high-quality ($\sim$$0\farcs6 - 0\farcs8$) images with CFHT of 27
CNELGs.  One galaxy shows clear evidence for a tidal tail; the others
are not unambiguously embedded in galactic disks.  Approximately 55\%
of the CNELGS have sizes consistent with local dwarfs of
small-to-intermediate sizes, while 45\% have sizes consistent with
large dwarfs or disks galaxies.  At least 4 CNELGs cannot harbor
substantial underlying disk material; they are low-luminosity galaxies
at the present epoch (M$_{\rm B} > -18$).  Conversely, 15 
are not blue enough to fade to low-luminosity dwarfs (${\rm
M_B} > -15.2$).  The majority of the CNELGs are consistent with
progenitors of intermediate-luminosity dwarfs and low-luminosity
spiral galaxies with small disks.  CNELGs are a heterogeneous
progenitor population with significant fractions (up to 44\%) capable
of fading into today's faint dwarfs (${\rm M_B} > -15.2$), while 15 to
85\% may only experience an apparently extremely compact CNELG phase
at intermediate redshift but remain more luminous galaxies at the
present epoch.

\end{abstract}


\section{Introduction}

Cosmological lookback time is one of the most powerful tools available
to study galaxy evolution.  To see the phases at which galaxies build
their stellar populations rapidly, we look back in time to find
dramatic starbursts.  However, we must learn to identify the
present-day galaxy population of which these starbursts are the
progenitors.  In this paper, we work toward a broad understanding of
the role of rapid star formation in the evolution of galaxies by
investigating the nature of some of the most extreme starbursting
galaxies at intermediate redshift, the compact narrow emission-line
galaxies (CNELGs).  Although the CNELGs are rare in the
intermediate-redshift universe \citep[$\sim$$7.5 \times 10^{-5}$
Mpc$^{-3}$;][]{Koo94}, starburst timescales are much less than a
Hubble time.  Thus, starbursting galaxies may have a relatively low
space density even if most galaxies go through such a phase.

Originally defined as stellar in ground-based images of moderate
resolution ($\sim$$1\farcs4 \pm 0\farcs05$ FWHM), CNELGs are
intrinsically luminous star-forming galaxies (${\rm -22 \lesssim M_B
\lesssim -18}$) with extremely small half-light radii (${\rm 1
\lesssim R_e \lesssim 5\ kpc}$) and narrow emission line widths in
integrated spectra \citep[$28 \leq \sigma \leq 126$ km
s$^{-1}$;][]{Koo95, Guz96}.  The small sizes and narrow emission lines
of these galaxies led \citet{Koo95} to identify the CNELGs as possible
starbursting objects that will fade by many magnitudes to become
present-day low-luminosity dwarf elliptical (dE) galaxies.  However,
these authors also note an alternative explanation: that CNELGs are
starbursts embedded in more massive galaxies, surrounded by an old
stellar population.  The two CNELGs with detailed abundance
measurements in the literature are more metal-enriched in their gas
phase than the mean stellar metallicities of dwarf galaxies
\citep{Kob99}, potentially supporting this hypothesis as we consider
further in \S3.2.1.  In addition, kinematic linewidths can
underrepresent the masses of compact galaxies
\citep{Kob00,Bar01,Pis01}.  If CNELGs are starbursts embedded in more
massive galaxies, they may be building stellar mass in central
components such as bulges or cores forming {\it in situ}
\citep{Kob99,Ham01}.  They may represent the formation of a separate
class of bulges known as ``pseudobulges'' or ``exponential bulges''
that have properties consistent with {\it in situ} formation from the
infall of disk gas and subsequent central formation
\citep[e.g.,][]{Car99, Kor04}.

High-resolution images of the CNELGs with the {\it Hubble Space
Telescope} reveal lumpy high surface brightness structures within the
starbursts in many cases \citep{Guz98}.  However, the lack of
sensitivity to lower surface brightness features could hamper efforts
to find older stellar populations in which CNELGs might be embedded.

Here, we present deep ground-based imaging designed to observe the low
surface brightness flux distribution of CNELGs.  To interpret the
data, the approach we take is to compare the properties of CNELGs to
the various specific models of the nature of CNELGs, focusing on one in
which they are dwarf elliptical galaxies undergoing a period of rapid
star formation and one in which they are starbursts in the centers of
galactic disks.  In \S 2, we describe the CNELG sample and the data
reduction.  \S 3 is a detailed comparison of the CNELG surface
brightness distributions to starbursting dwarfs, disk galaxies,
elliptical galaxies, and E$+$A galaxies.  In \S 4 and \S 5, we discuss
the results and list our conclusions.  We adopt a cosmology in which
$\Omega_\Lambda = 0.7$, $\Omega_{\rm m} = 0.3$, and H$_0 =
70$~km~s$^{-1}$~Mpc$^{-1}$.

\section{The Sample and Data Reduction}

We study a subset of CNELGs that have existing Keck spectroscopy in
the literature.  The parent sample originates from the photographic
plate study of \citet{Kron80}, including the fields SA 57 and SA 68
observed to 50\% completeness depths of $B_J\sim23$ and $R_F\sim22$
\citep[see][]{Kron80,Koo86,Koo88,Munn97}.  We reobserve the fields in
the $R$ band for 6-7 hours in queue mode with the 12k camera at the
Canada-France-Hawaii Telescope (CFHT) during optimal seeing conditions
\citep{Mar02}.  The new data are roughly ten times deeper, with
roughly half the seeing disk of the original plate observations.

We exclude objects from the analysis that are too close the edges of
the detector, galaxies that are next to very bright stars, and one
unresolved galaxy with low-quality spectroscopy that would be much too
intrinsically luminous at its tentative redshift.  The final sample of
observed galaxies includes 27 CNELGs and 3 additional compact
galaxies.  Of the 27 CNELGs, 19 have high-quality redshift
identifications, 5 have one-line redshifts, and 3 have low-quality
redshifts.

The deep $R$-band images of the CNELGs are dithered pointings of
600-second duration spread over several nights.  We list the total
exposure times of high-quality data for each object in
Table~\ref{tab:CNELGs}. The seeing FWHM of the final combined images
ranges from $\sim0\farcs6$ to $0\farcs8$.  The preliminary data
reduction was conducted with the Elixir pipeline processing software
at CFHT \citep{Mag04}.  Nominally, the flat fields allow photometric
accuracy to $\sim$2\%.  The photometric solutions provided by Elixir
were verified by comparing the new results with previous observations
of these objects.  Further, for SA57, we re-computed the photometric
solutions by comparing with $R$-band data in \citet{Maj94}.  While
there was good agreement with the Elixir results, we use the revised
calibration for the object photometry we present for this field.  Our
post-processing with the MSCRED package in IRAF\footnote{IRAF is
distributed by the National Optical Astronomy Observatories.} includes
coordinate solutions, image projection, and combination.  Finally, we
subtract the background using a large mesh size with SExtractor
\citep{Ber96}.  The typical 1-$\sigma$ surface brightness limits for
the combined images are $\sim$28 mag arcsec$^{-2}$.

Table 1 lists the basic properties of the sample. We use the measured
m$_R$ from our data, but construct rest-frame $B-V$ colors and M$_B$
from a compilation of the photographic $B_J$ and $R_F$ magnitudes and
spectroscopic redshifts: from the Munn et al. (1997) study for the
three ``compact galaxies''; from the series of papers on deep QSO
surveys in SA 57 and SA 68 (Koo, Kron, \& Kudworth 1986; Koo \& Kron
1988; Trevese et al. 1989, 1994; and Bershady et al. 1998) for most of
the CNELGs in SA57; and from previously-unpublished catalogs related
to the Munn et al. survey for the remainder. We use the
transformations from Majewski (1992) to convert {\it observed}
$B_J-R_F$ to {\it observed} $B-V$, noting that these transformations
are strictly relevant to stellar spectral-energy-distribution (SEDs)
in the rest-frame.  Using galaxy SED templates from Coleman, Wu \&
Weedman (1980) and a starburst template from Bruzual \& Charlot
(2003), we interpolate between templates to match the observed-frame
$B-V$ colors and use the interpolated templates to derive rest-frame
$B-V$ colors and M$_B$. A comparison between rest-frame $B-V$
calculated here and in Guzman et al. (1998) yields the same mean
values for 8 objects, but a significant dispersion of 0.1
mag.\footnote{The Guzman et al. study uses the same redshifts, but
include the photographic $U$ and $I_N$ bands, and in some cases this
was augmented by HST/WFPC-2 $V_{606}-I_{814}$ colors, in their
analysis.}  We adopt this as an estimate of the uncertainty in
deriving rest-frame colors, which comes into play when we discuss the
possible fading of these sources in \S 3.1.1.

\subsection{Surface Brightness Distributions of the Compact Narrow Emission
Line Galaxies}

\begin{figure}
\plotone{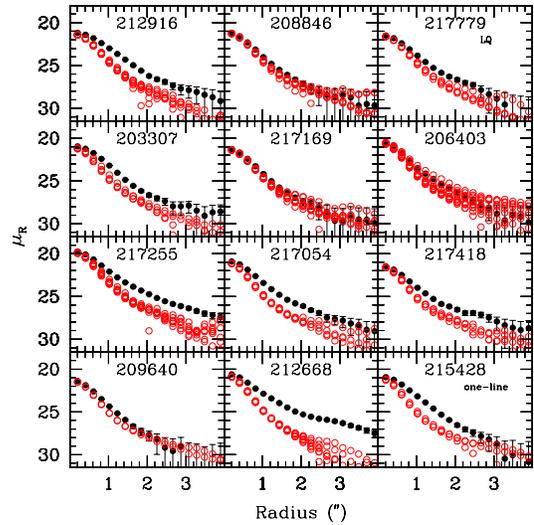}
\caption{$R$-band circular aperture surface brightness profiles of the
CNELGs.  The open circles show the point-spread function from profiles
of nearby stars scaled to the central surface brightness of the
CNELG. Labels are as follows: {\it oneline}: one-line spectroscopic
redshifts, {\it LQ}: low-quality spectroscopic data, {\it extended}:
extended source in original imaging data (not a CNELG).}
\label{fig:profiles}
\end{figure}

\begin{figure}
\protect \setcounter{figure}{0}
\plotone{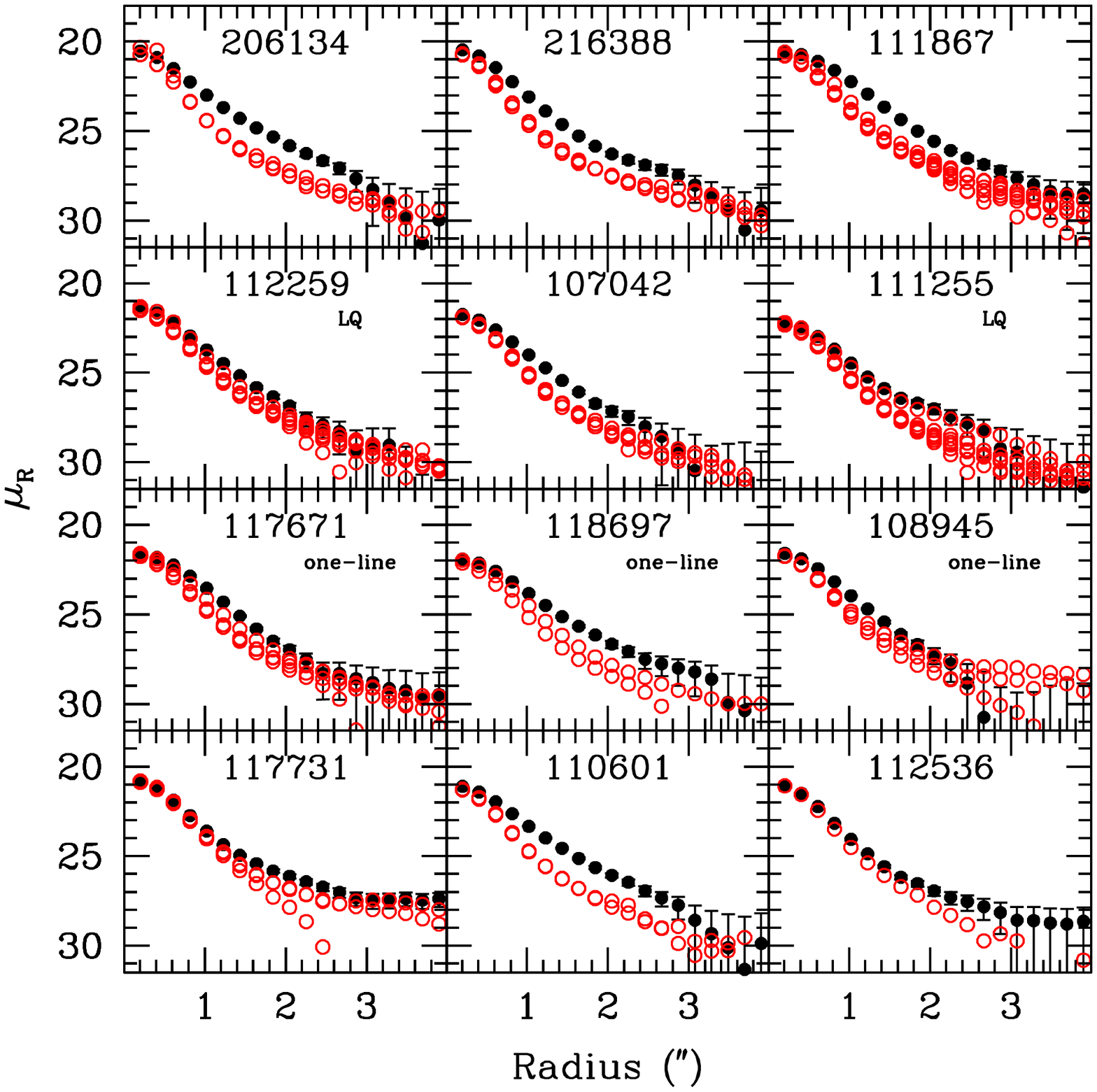}
\caption{Continued.}
\end{figure}

\begin{figure}
\protect \setcounter{figure}{0}
\plotone{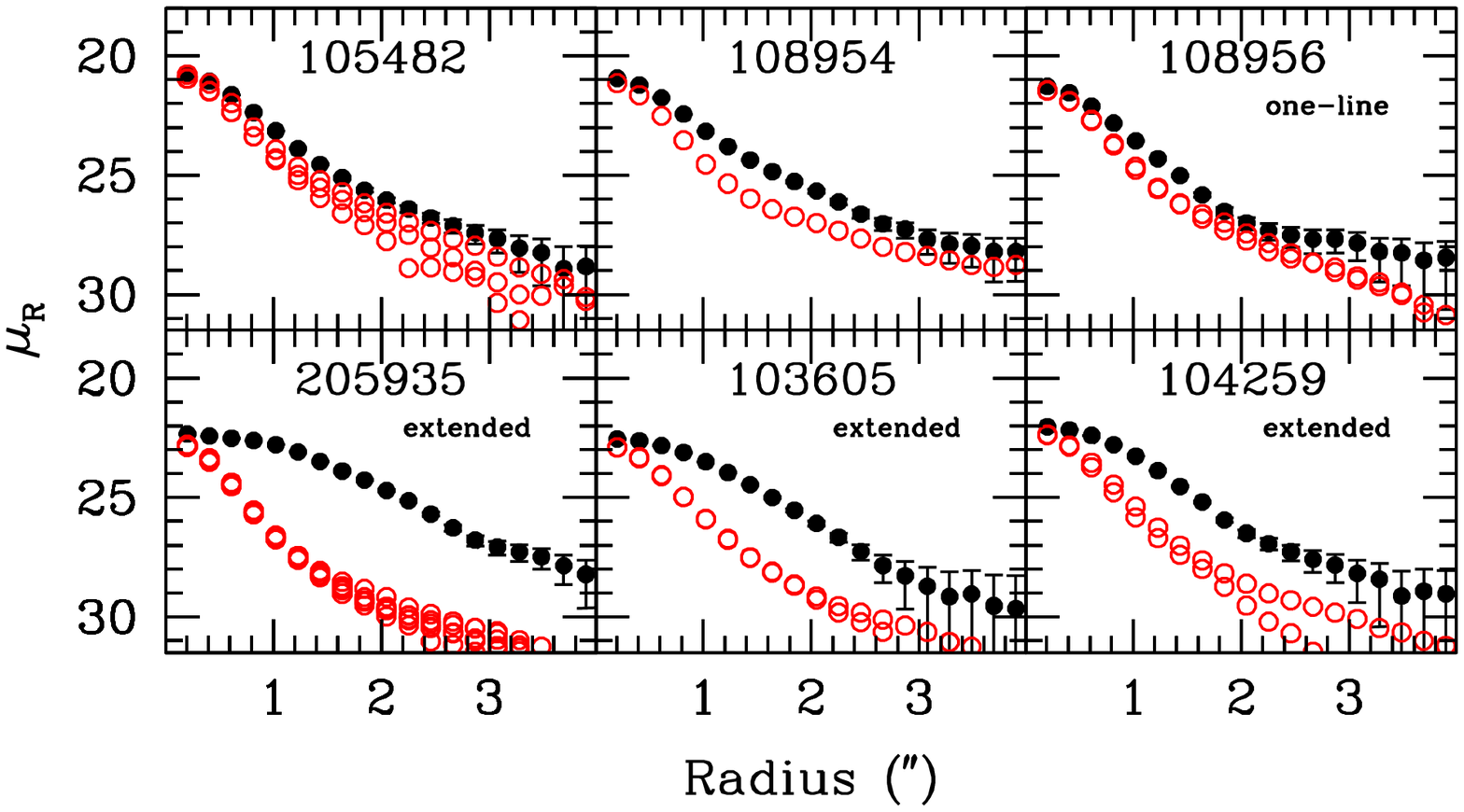}
\caption{Continued.}
\end{figure}

Fig.~\ref{fig:profiles} shows the surface brightness profiles of the
final sample ({\it filled circles}) based on circular aperture
photometry with SExtractor; the smaller open circles are the
profile(s) of a nearby star(s) scaled to the central surface
brightness of the CNELG.  Several CNELGs are unresolved or very nearly
unresolved in the images.  As we demonstrate below, most of these
unresolved galaxies are probably intrinsically small objects.
Fig.~\ref{fig:images} shows contour plots of the CNELGs.  The CNELGs
are clearly a heterogeneous population of objects.  Many of the CNELG
profiles resemble the stellar profiles in the inner regions,
suggesting that the inner parts of the galaxies are largely
unresolved.  Therefore, the sensitivity of our imaging observations to
detect underlying disks relies on the analysis of the faint outskirts
of the galaxies, where the depth of our observation must be combined
with careful data reduction and modeling to determine the types of
extended flux to which we are sensitive.

Background galaxies have a large filling factor in deep images of the
universe.  A superposition of a CNELG on a low surface brightness
background (or foreground) galaxy could lead to the misidentification
of structure in the outskirts of the CNELG.  We use a simple Monte
Carlo simulation to explore the probability that this effect is
significant.  The superposition would not be evident in even
high-quality spectra because the high surface brightness CNELG
dominates the flux.  We add 16 flux-boosted, artificially-redshifted
versions of NGC 205 (see \S~3.1 below) to randomly drawn positions in
the images and count the objects that deviate from the profile of a
simulation added to blank sky.  We do not count obvious superpositions
where the ``artificial CNELG'' (NGC 205) is separated from an adjacent
but distinct object; we count only those that we would likely mistake
for underlying structure.  For 480 simulated fake CNELGs, we find an
average rate of 23\% for superpositions, where 15\% are obvious
superpositions that would probably never have been identified as
CNELGs.  Of the remaining CNELG simulations, 8.6\% are accidental
superpositions that would not have been obvious.  Thus, our analysis
must allow for the fact that on average we expect that $2.3 \pm 1$ of
the 27 CNELGs are superpositions on extended, low surface brightness
background galaxies.

\begin{figure}
\plotone{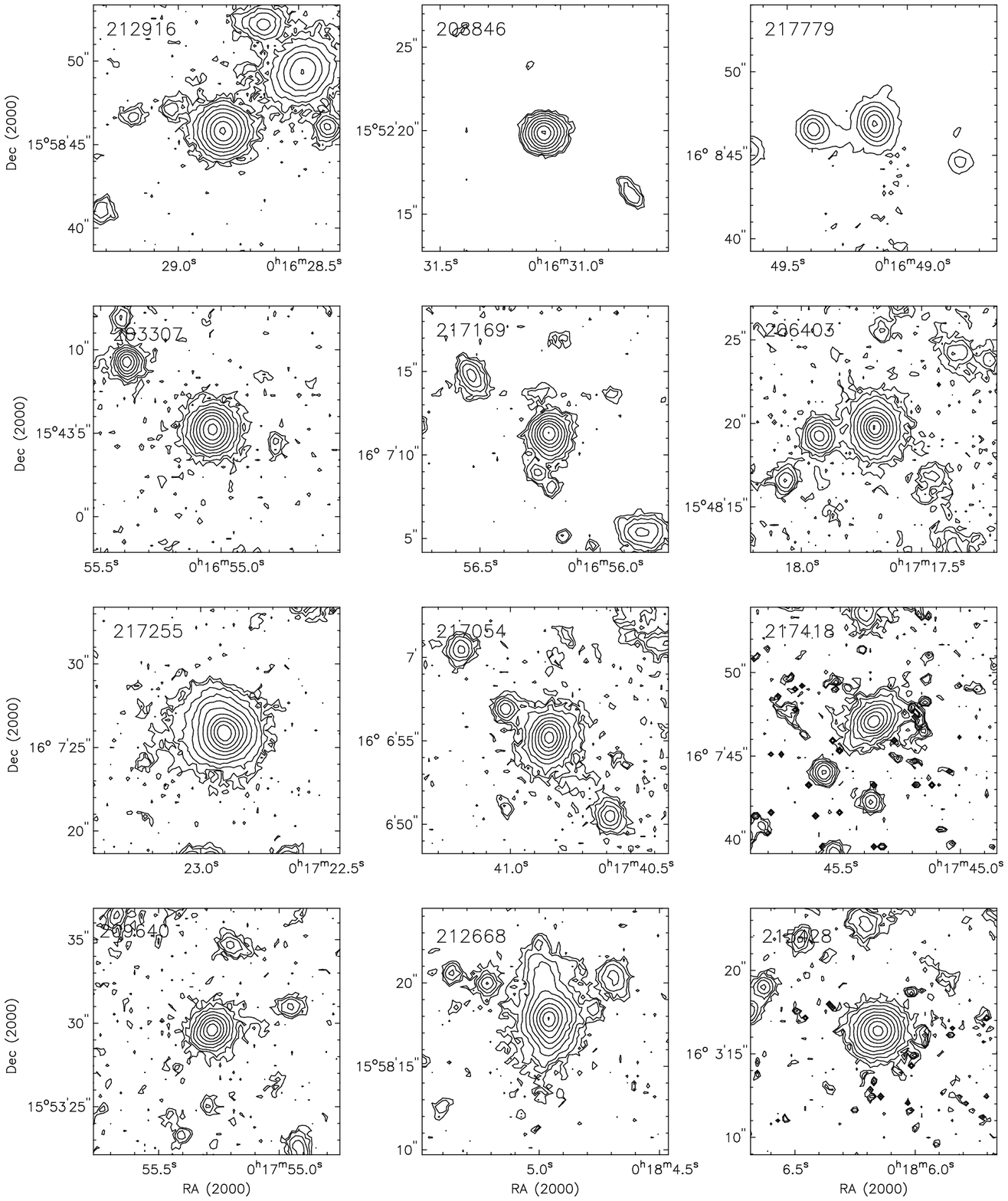}
\caption{Logarithmic contour plots of the $R$-band images of CNELGs.
The lowest contour corresponds to 27.3 mag arcsec$^{-2}$
for all except 107042, 108945, 111255, 112259, 117731, 216388 (26.5
mag arcsec$^{-2}$), 110601, 117671 (25.8 mag arcsec$^{-2}$),
and 217779 (25.3 mag arcsec${-2}$).  In all cases, the
contours are separated by 0.75 mag arcsec$^{-2}$.}
\label{fig:images}
\end{figure}

\begin{figure}
\protect \setcounter{figure}{1}
\plotone{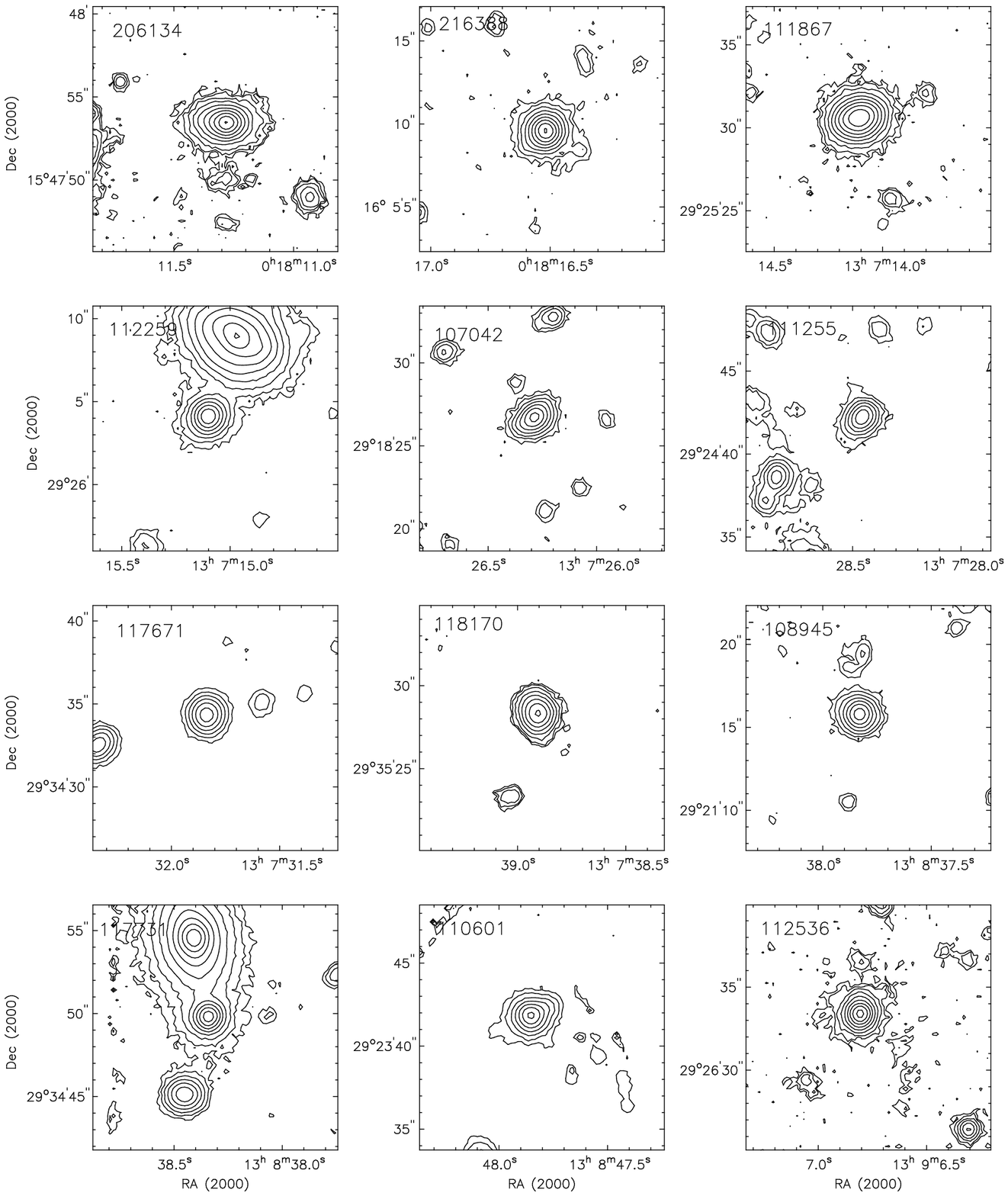}
\caption{Continued.}
\end{figure}

\begin{figure}
\protect \setcounter{figure}{1}
\plotone{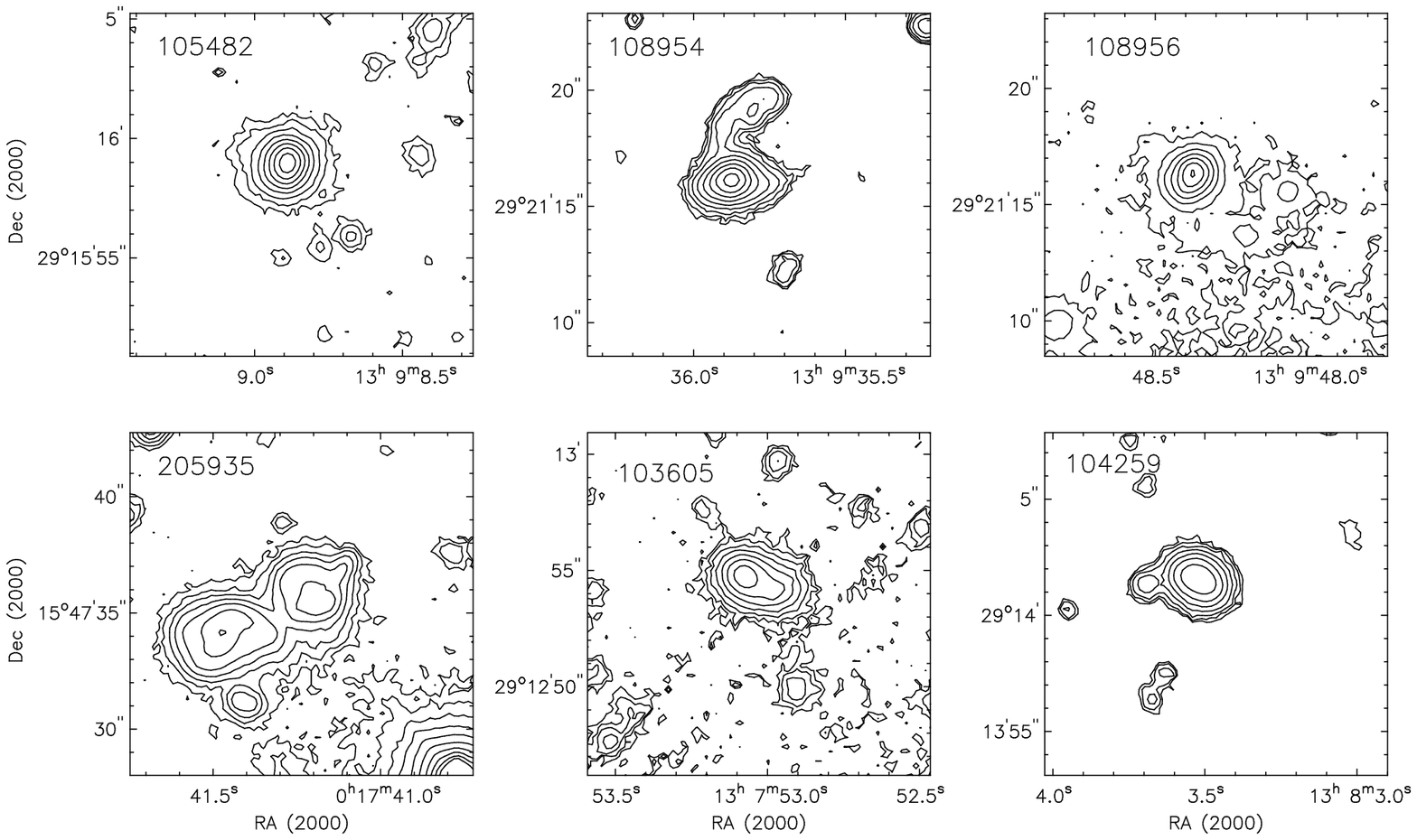}
\caption{Continued.}
\end{figure}

\section{Scenarios for the Nature of the CNELGs}

The defining characteristics of CNELGs are their extremely compact
sizes yet large blue luminosities, blue optical colors, and strong and
narrow line emission.  Studies of these objects and of similar
luminous compact blue galaxies reveal a heterogeneous population of
objects including some that probably fade to become dwarfs and some
star forming disk galaxies that fade to become luminous but
sub-L$^{\star}$ systems, just below the ``knee'' in the luminosity
function.  They exhibit a range of emission linewidths, a range of
stellar masses based on modeling of near-infrared fluxes, and a range
of morphologies
\citep{Koo94,Koo95,Guz96,Phil97,Guz97,Guz98,Ham01,Guz03}.  Here, we
scrutinize the surface brightness profiles of the CNELG population
from the SA57 and SA68 fields, comparing their surface brightness
profiles and colors to models based on the hypotheses that some CNELGs
are bursting dwarf galaxies, starbursts in the center of disk galaxies
(possibly galactic bulges in formation), the cores of elliptical
galaxies in formation, and/or the progenitors of E+A galaxies, which 
are galaxies with strong post-starburst spectra.

\subsection{The bursting dwarf hypothesis}

The bursting dwarf hypothesis posits that the luminous, compact
star-forming galaxies observed at intermediate redshift are the
progenitors to local dwarf galaxies.  In this scenario, the distant
bursting dwarfs deplete their gas reservoirs in a violent burst of
star formation, fading to become the dwarf elliptical galaxies (dEs)
observed in abundance in the local universe.  This interpretation of
the CNELGs is compelling because their emission linewidths and sizes
are appropriate for dwarfs and not for more luminous galaxies
\citep[e.g.,][]{Guz96}.  Here, we begin with the model that CNELGs
actually are dEs during a bursting phase and test the consistency of
that picture.

Formed stellar populations generally cannot shrink in size with time.
Thus, if CNELGs are starbursting dwarf galaxies that ultimately fade
to become dwarfs in the local universe, the sizes of CNELGs are
probably similar to or smaller than their counterpart dwarfs in the
local universe.  Supporting this conclusion, the CNELGs themselves are
{\it observed} to have radial color gradients (Guzman et al. 1998),
typically with blue cores.  If CNELGs fade to dEs, the aging stellar
populations in local dEs do not contradict the conclusion that they do
not shrink with time.  In particular, dEs exhibit only mild systematic
gradients in color or stellar population \citep[e.g.,][]{vader88,
jerjen2000, barazza2003, van04}; if anything, most are slightly
($\sim$0.1 mag.)  redder in their centers than their outskirts.
Therefore, we assume that CNELGs are smaller than or equal to the
sizes of the objects to which they ultimately fade.

\begin{figure}
\plotone{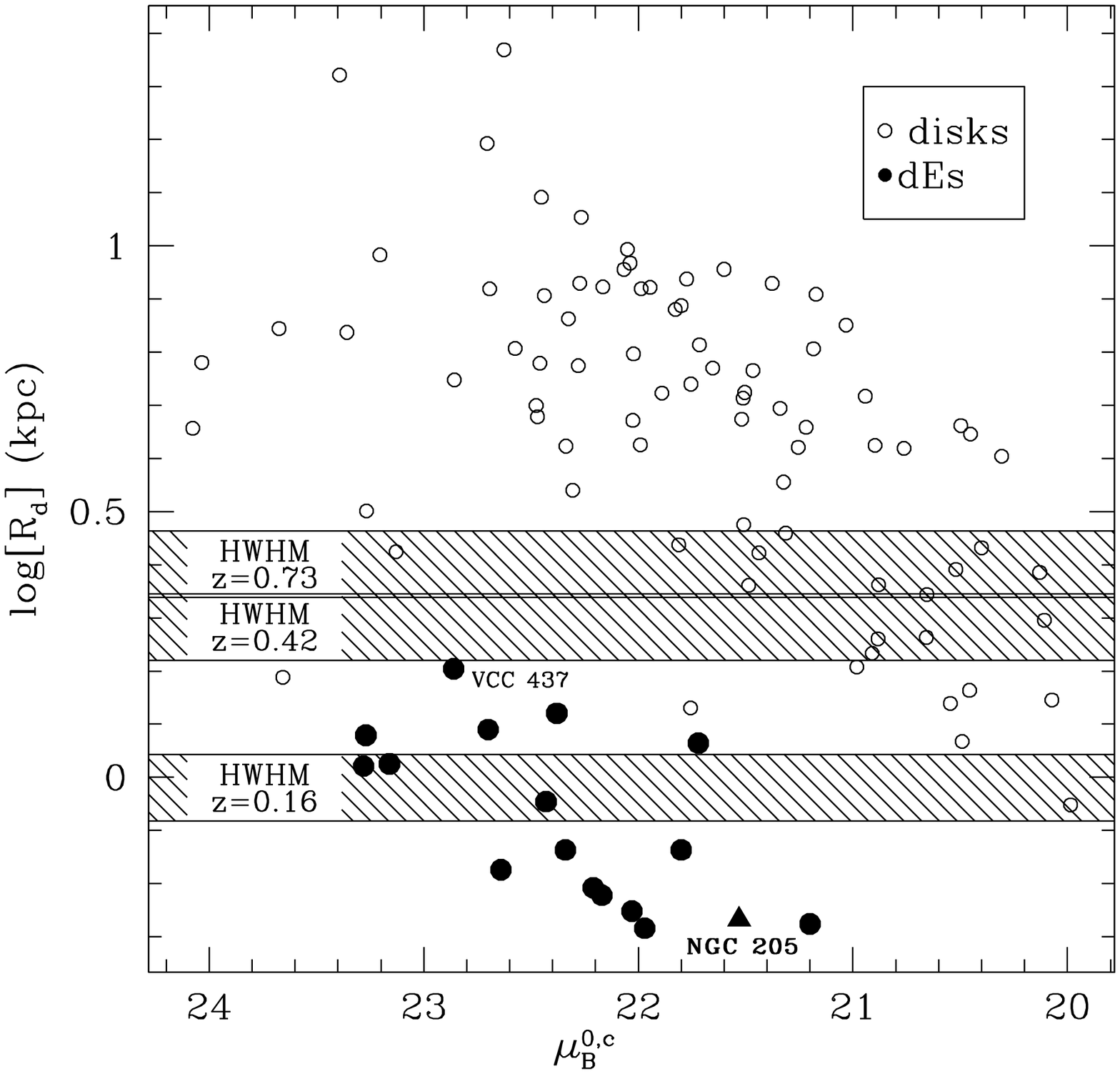}
\caption{Structural parameters for Virgo dEs ({\it filled circles})
from van Zee, Barton \& Skillman (2004), for NGC 205 ({\it filled
triangle}), and for local disks ({\it open circles}) from de Jong
(1996).  We plot the log of the half-light radii of the galaxies as a
function of the disk rest-frame $B$-band central surface
brightness. We shade the half-width at half maximum (HWHM) of the
typical stellar point-spread functions for our data (range
0\farcs6~--~0\farcs8) at different redshifts.  Low-luminosity dwarfs
would be barely resolved or unresolved, except at the lowest
redshifts; the larger dwarfs and most disks would be resolved.}
\label{fig:ngc205}
\end{figure}

To test the hypothesis that CNELGs are the progenitors of dwarfs, we
compare their sizes directly to examples of both compact and more
extended local dEs.  We use this analysis to argue that objects that
are larger than dE's do not fade to become dE's by the present epoch,
taking examples from a wide range of small and large dE's.  The
fiducial example of a small dwarf CNELG analog is NGC 205, a dwarf
companion to M31.  The Virgo cluster holds a more varied population of
dEs including galaxies that are much larger than NGC 205.
Fig.~\ref{fig:ngc205} shows the central surface brightnesses and scale
lengths of NGC 205,\footnote{ Observations of NGC 205 were obtained
with the MOSAIC camera on the WIYN 0.9-meter telescope on 2003
December 19.  The observations consisted of three 300 s exposures in
the B-band; the observations were dithered substantially to fill in
the interstitial gaps.  While the night was non-photometric,
photometric calibration was obtained by comparison with the integrated
magnitudes tabulated in Mateo (1998).} and a set of Virgo dEs
\citep{van04}.  The shaded regions show the half-width at half maximum
(HWHM) of the range of seeing disks in the CFHT data at different
redshifts.  Although this figure serves as a guide to the sizes of
these objects, whether or not we can resolve these objects depends on
their luminosities and surface brightness profiles.  We explore these
questions in more detail below, using simulations added to the data.

Using a rest-frame B-band image of NGC 205 from the WIYN 0.9-meter
telescope and B-band images of elongated Virgo dEs from
\citet{van04}, we evaluate the bursting dwarf hypothesis by comparing
the surface brightness profiles of CNELGs to artificially redshifted
versions of dwarf galaxies.  To construct the model bursting dwarfs,
we boost the {\it integrated} flux of the model to the measured
$R$-band apparent magnitude of the CNELG with which we are comparing.
We re-compute the physical scale of the image at the redshift of the
CNELG and rebin the data to the $0\farcs205$ resolution the CFHT data.
We convolve with the point-spread function using a nearby star.  We
simulate the appropriate Poisson noise for the object flux and add the
object to the frames near the relevant CNELGs.  We measure surface
brightness profiles with a procedure that is identical to that used
for the CNELGs themselves.

\begin{figure}
\plotone{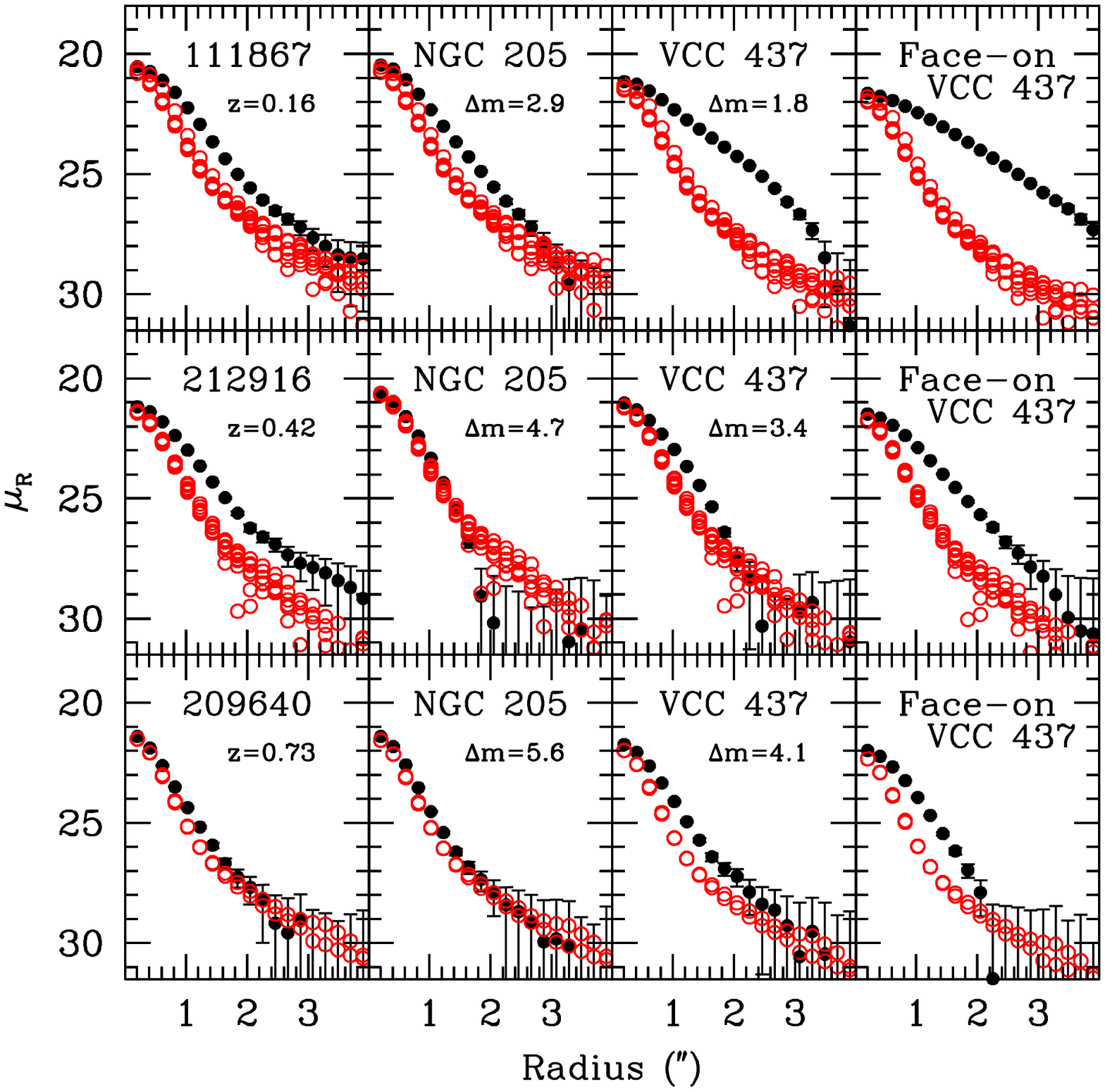}
\caption{Comparing compact CNELGs to models of artificially redshifted
dwarf galaxies.  In each row, we plot the circular aperture surface
brightness profile of a CNELG ({\it left, black}),
labeling the plot with the ID number and redshift.  The remaining
three panels, from left to right, are (1) an artificially redshifted
version of NGC 205 at the redshift of the CNELG with a magnitude boost
given by $\Delta$m, (2) an artificially redshifted VCC 437 --- the
largest dwarf in the van Zee, Barton \& Skillman (2004) sample of
elongated Virgo dEs, with a scale length of ${\rm R_d = 1.6}$~kpc ---
and (3) an artificially redshifted ``face-on'' model version of VCC
437 that is an exponential disk with the scale length and central
surface brightness of VCC 437.   In each panel we plot surface brightness
profiles of nearby stars scaled to the central surface brightness of
the object ({\it red}).  NGC 205 is only resolvable at the
lowest redshifts; most of the other Virgo dEs are barely resolved,
except those with long disk scale lengths.  In contrast, most known
galactic disks would be resolved.}
\label{fig:fakedwarfs}
\end{figure}

\begin{figure}
\plotone{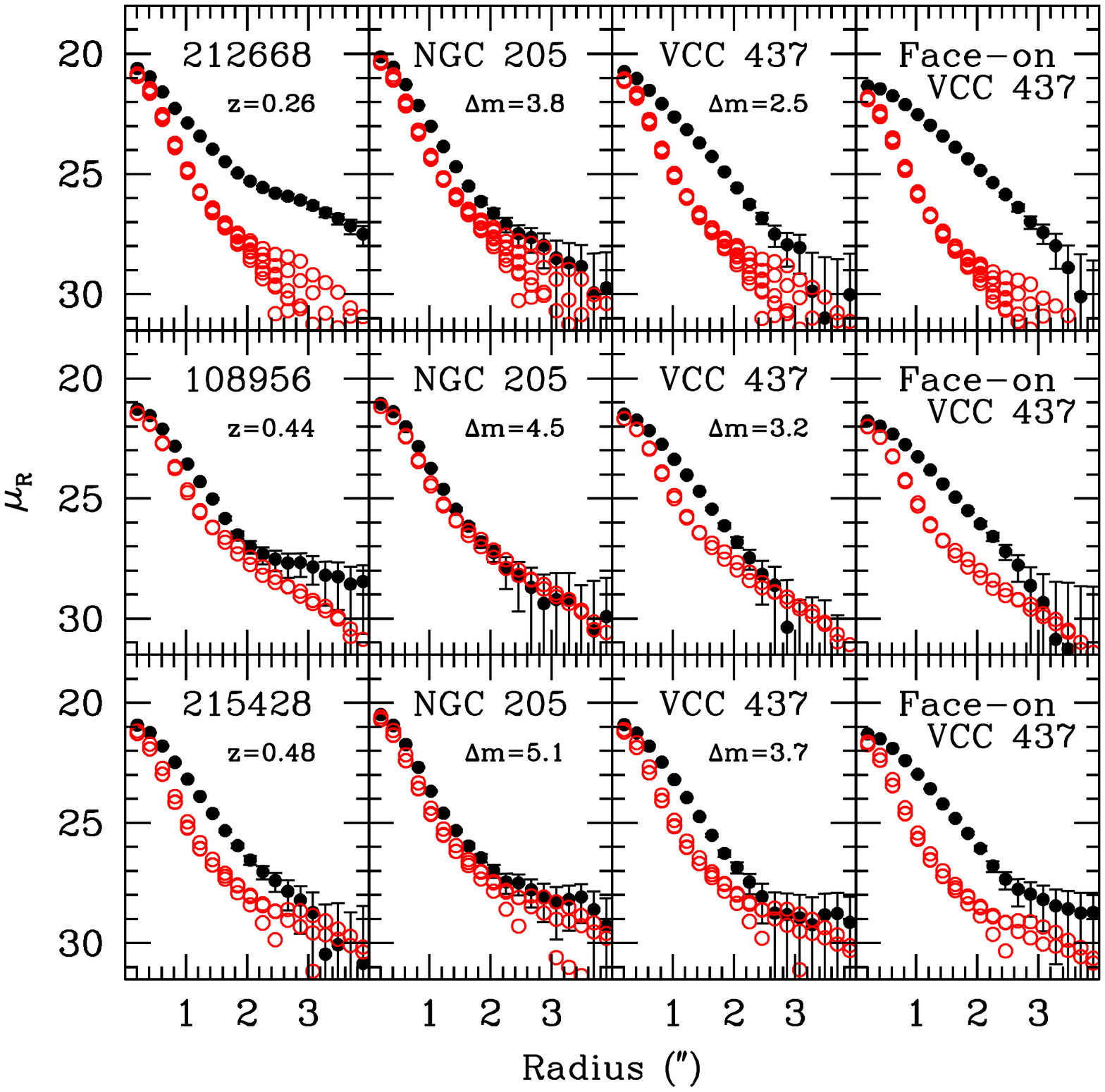}
\caption{Same as Fig.~\ref{fig:fakedwarfs}, but for more extended CNELGs.}
\label{fig:fakedwarfs2}
\end{figure}

Figures~\ref{fig:fakedwarfs} and \ref{fig:fakedwarfs2} show examples
of the resulting boosted-dwarf surface brightness profiles.  The red
points are nearby stars boosted to match the central surface
brightness of the CNELG or artificial dwarf.
Fig.~\ref{fig:fakedwarfs} includes examples at the highest and lowest
redshifts of the sample (111867 and 209640, respectively), and 212916
at the median redshift, $z=0.42$. The figures highlight the fact that
our discriminatory power varies as a function of redshift.  For
example, our data resolve NGC 205 at $z=0.16$ but it is essentially
unresolved at $z=0.73$.  However, we would be able to resolve VCC 437
at any redshift in our range.

In Fig.~\ref{fig:fakedwarfs}, 111867 and 209640 are similar to NGC
205, but the remaining galaxies are larger or have detected flux
extending to larger radii.  Fig~\ref{fig:fakedwarfs2} shows that
212668 is more extended than even our model version of the largest
Virgo dE from \citet{van04}, VCC 437, viewed as though it were a
face-on disk.  It has an exponential scale length of ${\rm R_d =
1.6}$~kpc that is large enough to fall within the size range of small
disk galaxies.  Similar figures for the remaining galaxies indicate
that of the CNELGs, 7/27 (26\%) are smaller than or similar in size to
NGC 205, 8/27 (30\%) are between NGC 205 and VCC 437 and thus
plausible typical dEs, 8/27 (30\%) are similar in spatial extent to
VCC 437 and thus consistent with a large dwarf or small spiral. 4/27
(15\%) are more extended than even the largest dE in the \citet{van04}
sample.  We list these broad classifications in Table~\ref{tab:CNELGs}
and use them to group the CNELGs hereafter.

To compute the true amount of (integrated) $B$-band luminosity boost
required to match the integrated $R$-band flux of the CNELG, we use
the interpolated template spectral energy distribution corresponding
to the rest-frame $B-V$ color of the CNELG to generate k-corrections
for the hypothetical bursting dwarf (see \S 2).  For NGC 205, the
required boosts range from 1.5 to 6.0 magnitudes.  The Virgo dEs fall
into a more reasonable range.  For the most luminous dE, VCC 1261, the
required burst luminosity boosts satisfy ${\rm 0.1 < \Delta {\rm m} < 4.6}$;
for the least luminous dE of the \citet{van04} sample, the
requirements are in the range ${\rm 2.1 < \Delta {\rm m} < 6.6}$. Below, we
examine the constraints on luminosity fading for individual CNELGs
available from their rest-frame colors.

\subsubsection{Fading Limits for the CNELGs}

\begin{figure}
\plotone{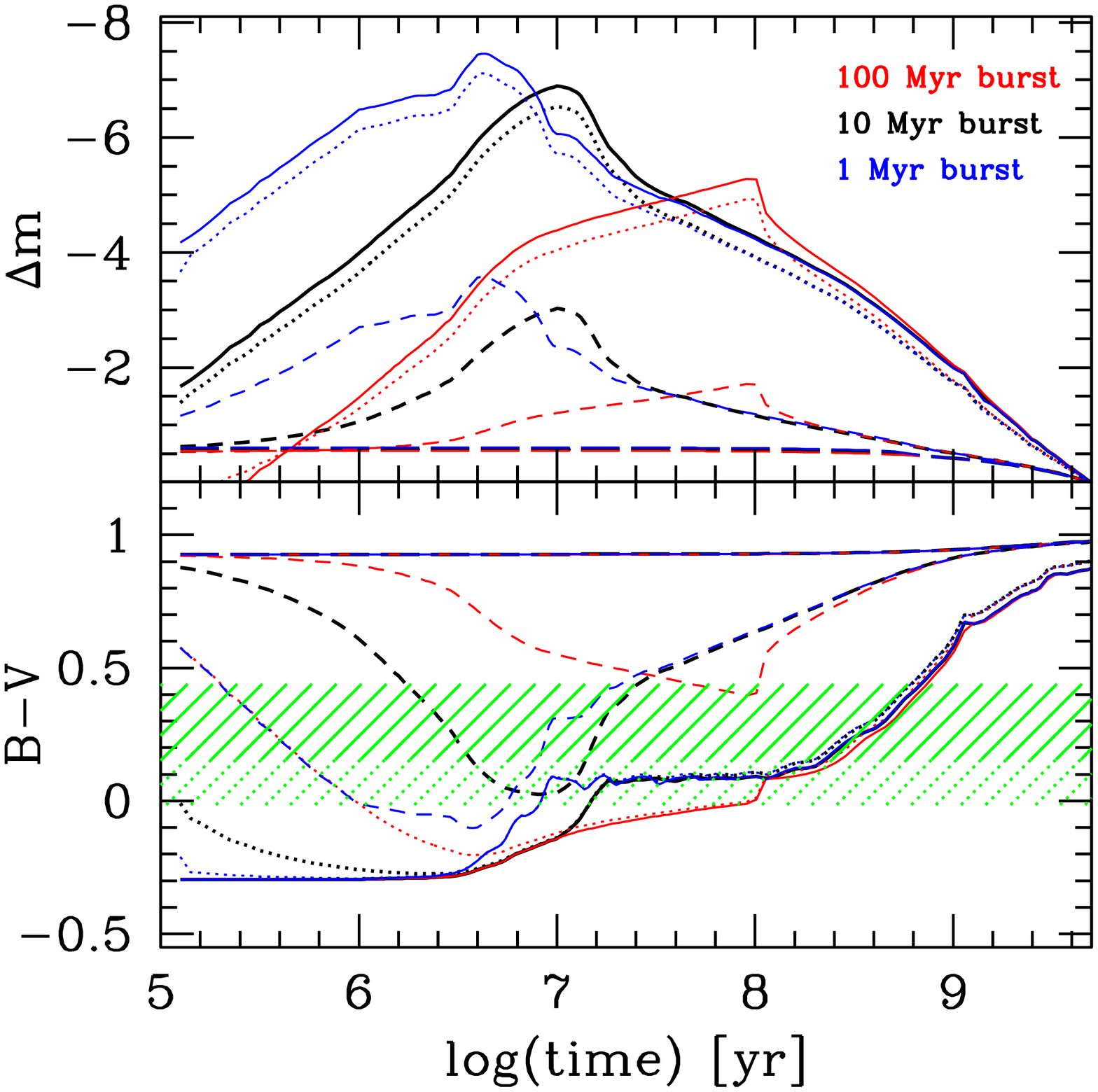}
\caption{Starburst colors and fading limits from \citet{Bru03} stellar
spectral synthesis models.  We plot the amount of fading ({\it top})
and the composite rest-frame $B-V$ color ({\it bottom}) for starbursts
with amplitudes of 0\% ({\it long-dashed line}), 1\% ({\it
short-dashed line}), 50\% ({\it dotted line}), and 100\% ({\it solid
line}) for starbursts with constant star formation rates for 1 Myr
({\it blue}), 10 Myr ({\it black}) and 100 Myr ({\it red}).  The
shaded region indicates the observed color range in our sample.  This
color range corresponds to a maximum fading of $\sim$3.5 magnitudes in
the 5 Gyr following the burst ($z=0.5$).}
\label{fig:bursts}
\end{figure}

Spectral synthesis models provide a means of estimating an upper limit
to the amount that CNELGs have faded to the present day.
Fig.~\ref{fig:bursts} shows the amount of fading and the composite
colors for \citet{Bru03} models with added starbursts. Adopting a
\citet{Chab03} stellar initial mass function and solar metallicity, we
assume two episodes of star formation that are bursts of equal
duration.  The second burst occurs 8 Gyr after the first with an
amplitude of 0\%, 1\%, 50\%, or 100\%.  On the plot, time is measured
since the start of the second burst. More rapid bursts reach higher
fractional peak luminosities.

\begin{figure}
\plottwo{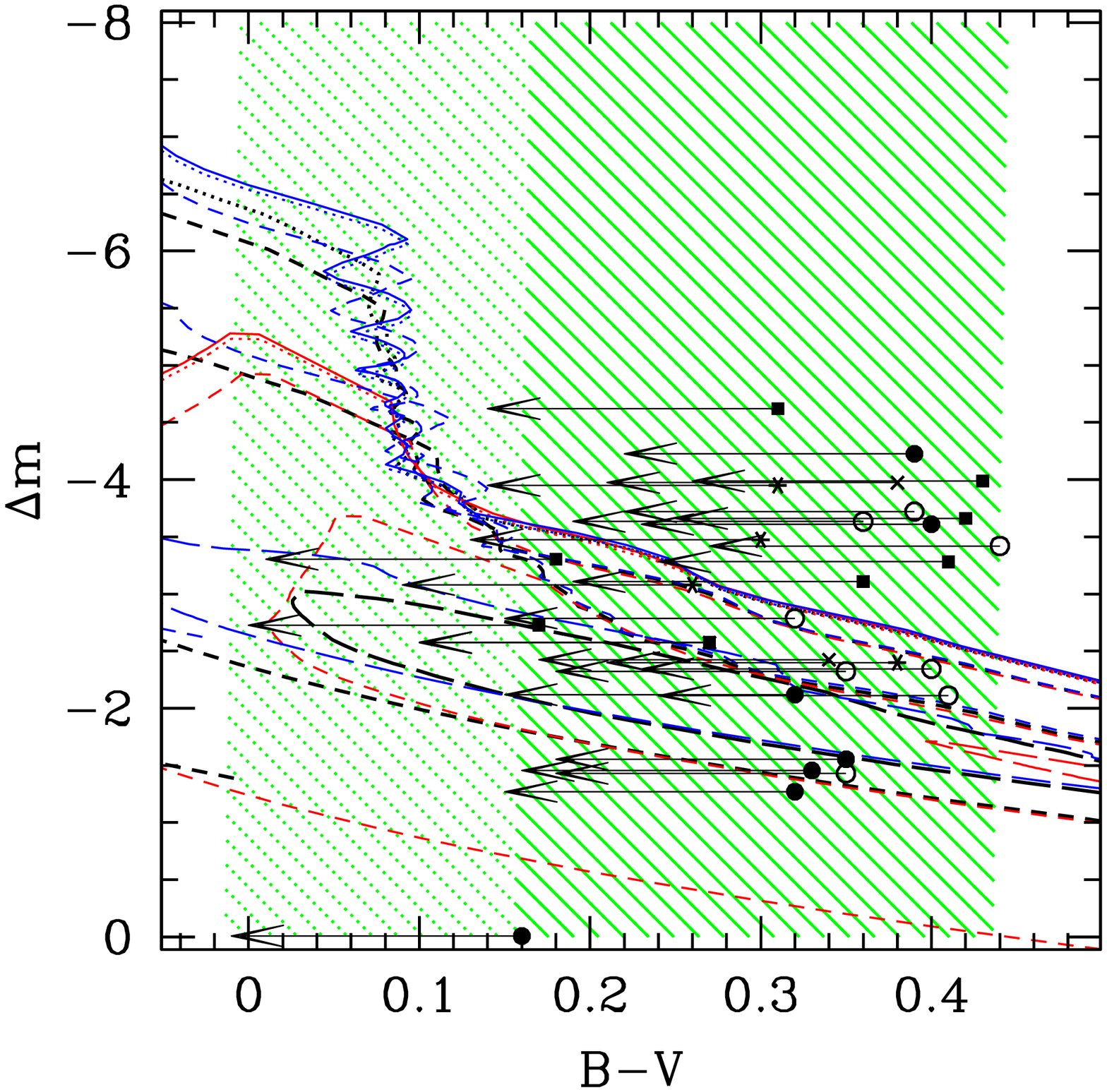}{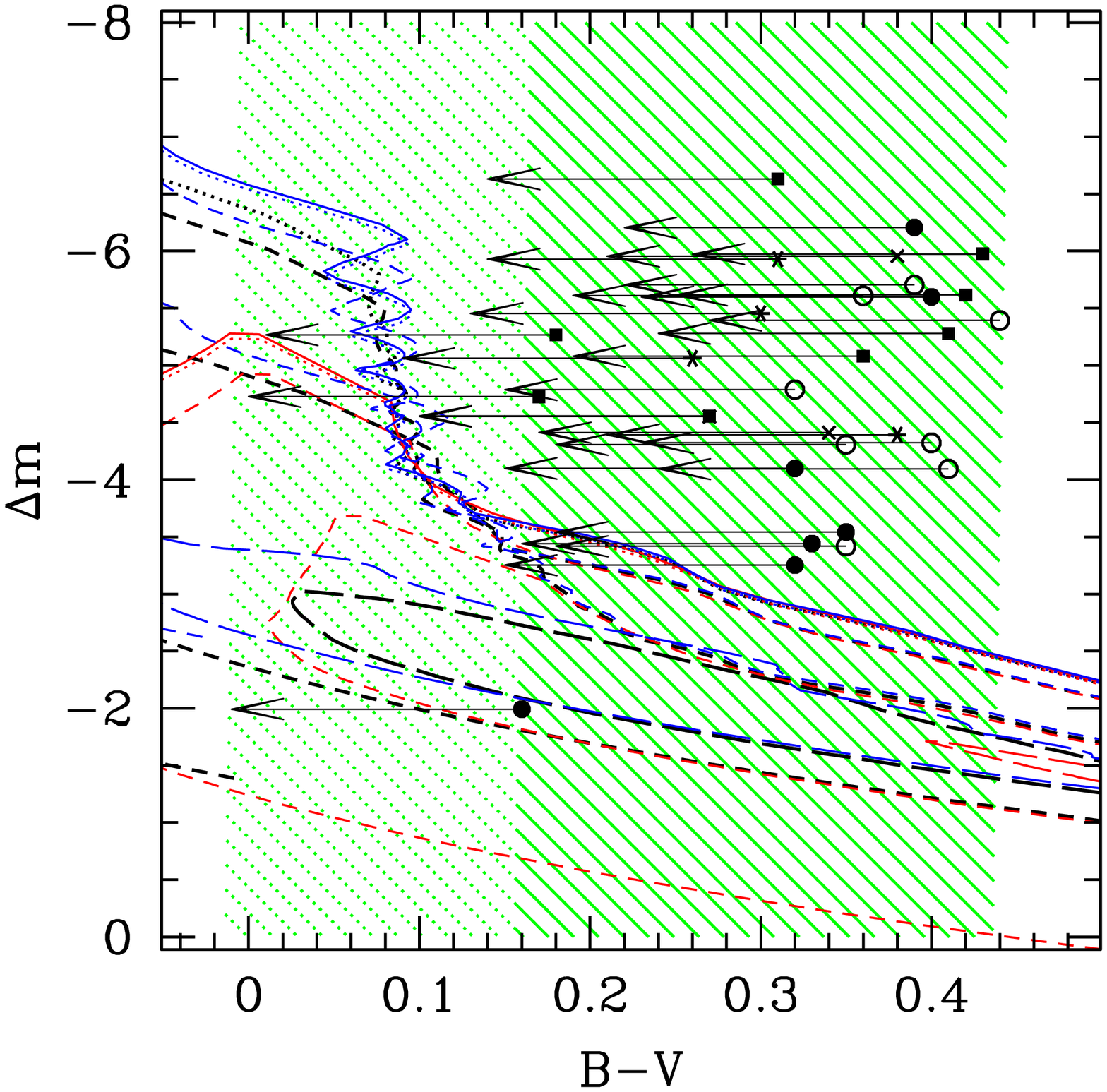}
\caption{Rest-frame color as a function of number of magnitudes the
burst will fade by the time 5 Gyr has elapsed since the burst.  The
lines are the models as in Fig~\ref{fig:bursts}.  The points show the
CNELGs that are NGC~205-like ({\it filled circles}), INTERMEDiate
between NGC~205 and VCC~437 ({\it open circles}), VCC~437-like ({\it
filled squares}), LARGE CNELGs ({\it asterisks}), and the additional
luminous compact blue galaxies ({\it crosses}).  We plot each point at
its observed rest-frame $B-V$ color on the x-axis vs. the number
of magnitudes the CNELG needs to fade to become the brightest dwarf in
the van Zee sample, VCC 1261 with M$_{\rm B} = -17.4$ ({\it left}), or
the number of magnitudes the CNELG needs to fade to become the
faintest dwarf in the van Zee sample, VCC 1743, with M$_{\rm B}=-15.2$
({\it right}).  The arrows indicate the rest-frame $B-V$ colors
assuming a reddening of $E(B-V)=0.17$.}
\label{fig:colormag}
\end{figure}

For global bursts of star formation, the burst timescales must be
$\gtrsim$ the dynamical timescales.  With central velocity dispersions
$25 \lesssim \sigma_0 \lesssim 50$ km s$^{-1}$ \citep{Geha03,van04b}
and disk scale lengths $0.5 \lesssim \alpha^{-1} \lesssim 1.6$ kpc
\citep[e.g.,][]{van04}, the dynamical timescales of the Virgo dEs are
always $\gg 10$ Myr and usually $\gtrsim 100$ Myr.  Thus, we consider
models in which the bursts last 10 Myr and 100 Myr as typical
timescales for global starbursts in a dE.  For comparison, estimated
dynamical timescales for CNELGs yield a mean of 20 Myr within the
half-light radius as observed at intermediate redshift.\footnote{
We use $t_{dyn}(R_e) = 1.014 R_e /
\sigma$, which comes from adopting $t_{dyn}(r) = \frac{\pi}{2} (G
M(r)/r^3)^{-1/2}$ \citep{BT87}, and the dynamical mass
formula $M = 3 (k/G) R_e \sigma^2$ \citep{Guz96} assuming $k =
1.6$.  The values of $R_e$ and $\sigma$ are from \citet{Koo94,Koo95}
and \citet{Guz96}.}
More localized bursts of star formation may last for shorter
amounts of time.  Thus, we include 1 Myr bursts to encompass this
possibility.  We note that abundance patterns in nearby dwarfs appear
to rule out star formation histories that involve extremely rapid
formation of $\sim$100\% of the stellar population: such a scenario
would result in alpha-enhanced dwarfs, while dEs typically have solar
or sub-solar [$\alpha$/Fe] ratios \citep{Geha03, van04}, indicating
that their dominant stellar population was not formed in a single
rapid starburst.  In any case, we find that the derived fading limits are
insensitive to choice of burst timescale in the explored range.

Our inability to correct for extinction accurately, on an
object-by-object basis, is the most important systematic error of this
analysis.  The average measured $E(B-V)$, derived for 7 CNELGs by
Kobulnicky \& Zaritsky (1999) and Hoyos et al. (2004), is 0.17.
However, if the CNELGs are starbursting dwarfs that resemble strongly
star-forming dwarfs in the nearby universe, their reddening may be
minimal.  Thus, we compute the expected fading of CNELGs assuming both
no dust and $E(B-V)=0.17$.

As Fig.~\ref{fig:bursts} shows, objects that will fade by $\gtrsim
3.5$ magnitudes ($\Delta {\rm m} \leq -3.5$) in the 5 Gyr after the burst
are bluer than the observed CNELGs for any of these scenarios, unless
the rest-frame colors are corrected for dust.  A correction of
$E(B-V)=0.17$ increases the allowed fading to 3-4 magnitudes in most
cases, and to $\geq 6$ magnitudes in the very bluest 3
CNELGs.\footnote{As noted in Table~1, the three bluest CNELGs are not
  necessarily the smallest.  One is consistent with the size of NGC
  205, but 2/3 are consistent only with the much larger VCC 437.}
Extending the amount of fading time to the lookback time for the
highest redshift of the sample allows $\sim$4 magnitudes of fading and
decreasing it to the lowest redshift decreases the allowed fading to
$\sim$3 magnitudes.  Reducing the assumed metallicity to 0.02
Z$_{\sun}$, which may be appropriate for the smallest dwarfs but is
not warranted by the small amount of existing data for CNELGs
\citep{Kob99}, decreases the fading allowed in 5 Gyr to $\sim$2.5
magnitudes.

Despite the effects of burst duration on the peak luminosity boost
allowed by a given burst of star formation, the relationship
between the total amount a starburst can fade and the color of the
galaxy is nearly independent of the timescale of the burst.  In
Fig.~\ref{fig:colormag}, we plot this relationship for the models
shown in Fig.~\ref{fig:bursts}.  The figures show rest-frame color as
a function of number of magnitudes the burst will fade by the time 5
Gyr has elapsed since burst.  The points show the CNELGs and the
additional luminous compact blue galaxies ({\it crosses}).  We plot
each point at its observed rest-frame $B-V$ color on the x-axis, and
at the number of magnitudes the CNELG needs to fade to become the
brightest dwarf in the van Zee sample, VCC 1261 with M$_{\rm B} =
-17.4$ ({\it left}), and the number of magnitudes the CNELG needs to
fade to become the faintest dwarf in the van Zee sample, VCC 1743,
with M$_{\rm B}=-15.2$ ({\it right}).  Arrows indicate the intrinsic
rest-frame $B-V$ if $E(B-V)=0.17$.  

\begin{figure}
\plottwo{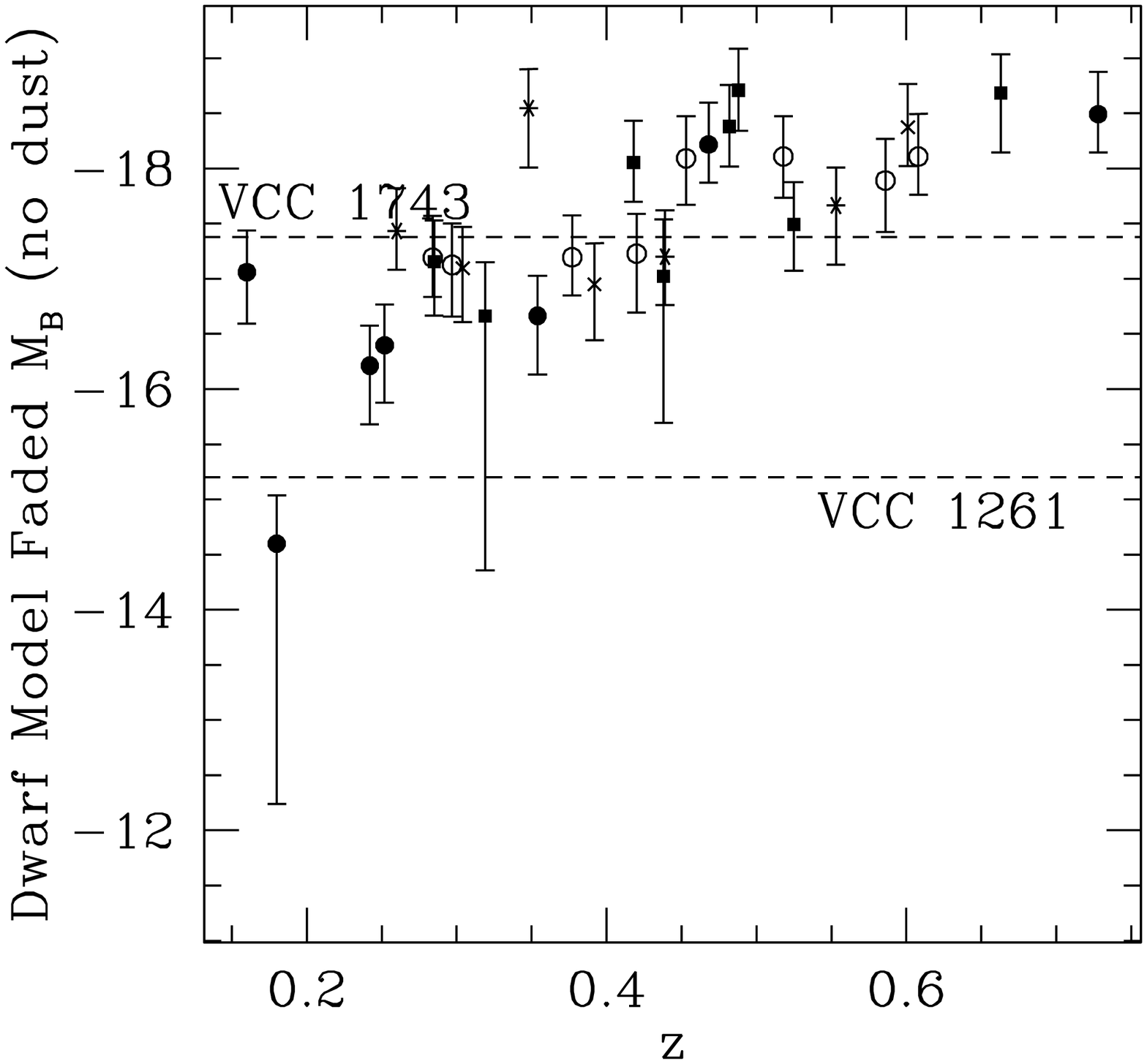}{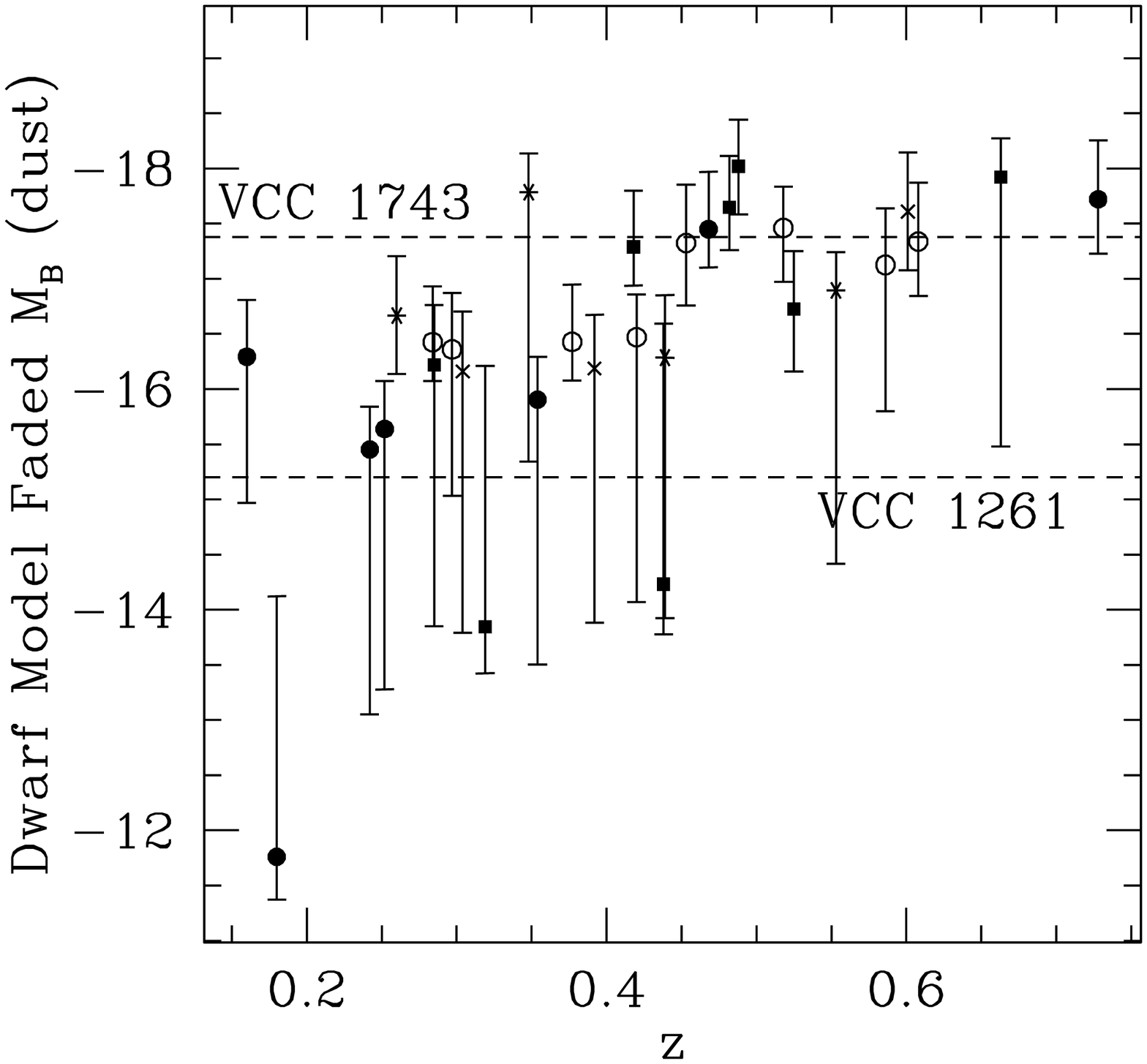}
\caption{The maximum present-day magnitudes (or the minimum present
day luminosities) of the CNELGs as a function of measured redshift,
computed from the maximum amount of fading allowed by the color and
lookback time for solar metallicity assuming no reddening ({\it left})
or $E(B-V)=0.17$ ({\it right}).  Error bars reflect an error in
rest-frame color, $(B-V)_0$, of 0.1 magnitudes.  Point types are as in
Fig~\ref{fig:colormag}. The horizontal dashed lines indicate the
luminosity range of the Virgo dE sample of \citet{van04}.}
\label{fig:fade}
\end{figure}

We list the true upper limit to the amount of fading allowed by this
analysis in Table~\ref{tab:CNELGs} calculated using the lookback time
to each object (set to 5 Gyr in Fig.~\ref{fig:colormag}), both with
and without reddening.  The upper limits correspond to the position on
the solid line for the color of the galaxy in Fig.~\ref{fig:colormag}
if the models were shifted each time to normalize at the lookback time
of the redshift of the CNELG.  For example, renormalizing the fading
dwarf to correspond to a fading time of 6.43 Gyr, the lookback time at
the highest redshift in our sample ($z=0.728$), instead of the 5 Gyr
used in Fig.~\ref{fig:colormag}, increases the allowed amount of
fading by 0.24 magnitudes.  In Figs.~\ref{fig:fade}a and b we plot the
minimum faded (present-day) magnitude of each CNELG as a function of
redshift with and without the reddening correction, respectively.  In
any scenario, all the galaxies can fade to become sub-L$^{\star}$
galaxies with M$_{\rm B} < -19$.  Without the reddening correction and
neglecting errors, 14/27 cannot fade to ${\rm M_B > -17.4}$ (the most
luminous dwarf in the van Zee sample) and 26 cannot to ${\rm M_B >
-15.2}$ (the faintest in the van Zee sample).  With the reddening
correction and neglecting errors, 7 cannot fade to ${\rm M_B > -17.4}$
and 24 cannot fade to ${\rm M_B > -15.2}$.  With the reddening
correction and the bluest color allowed by the estimated error, all
but one of the galaxies can fade to ${\rm M_B > -17.4}$ and 15 cannot
fade to ${\rm M_B > -15.2}$.

\begin{figure}
\plottwo{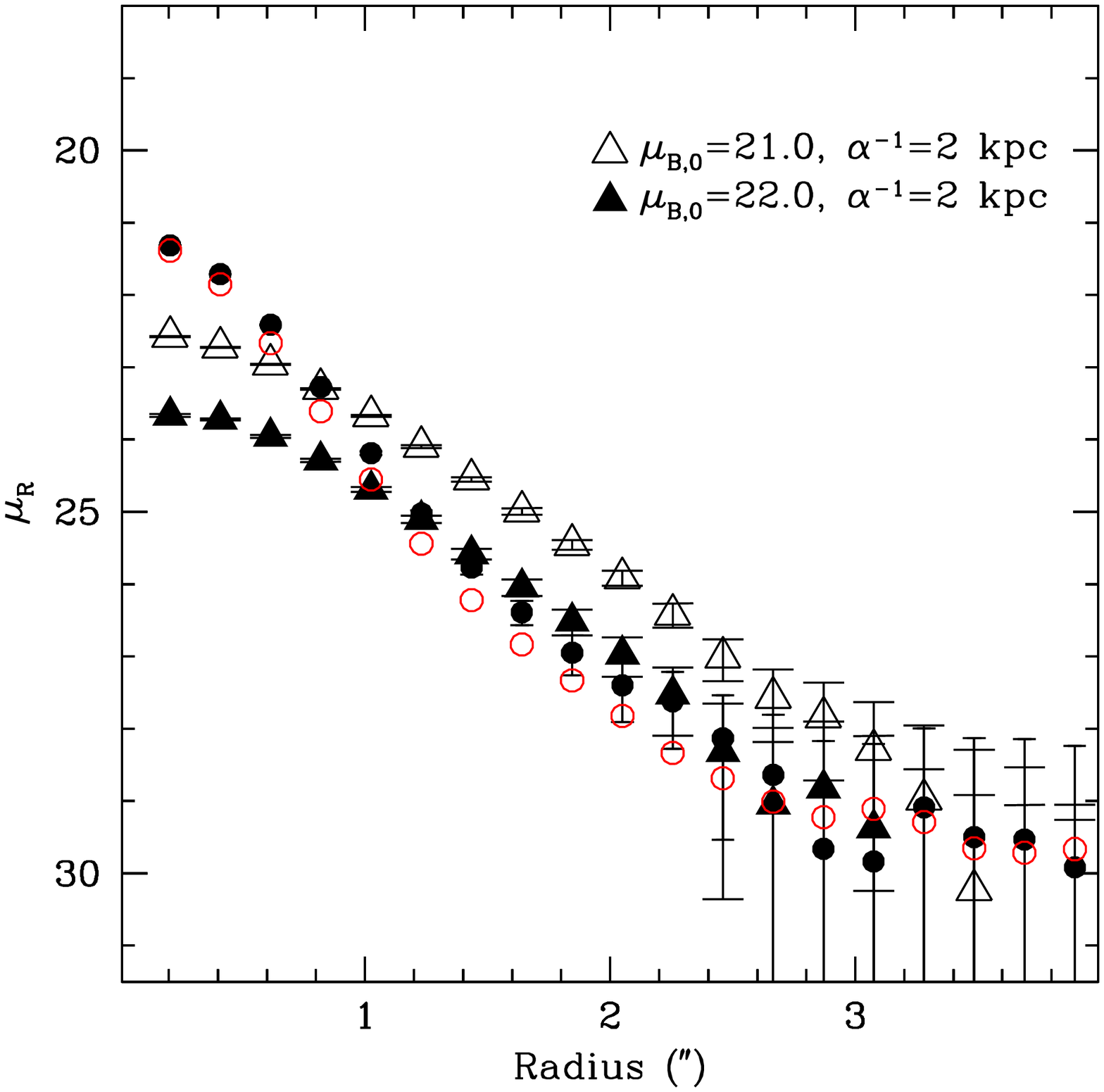}{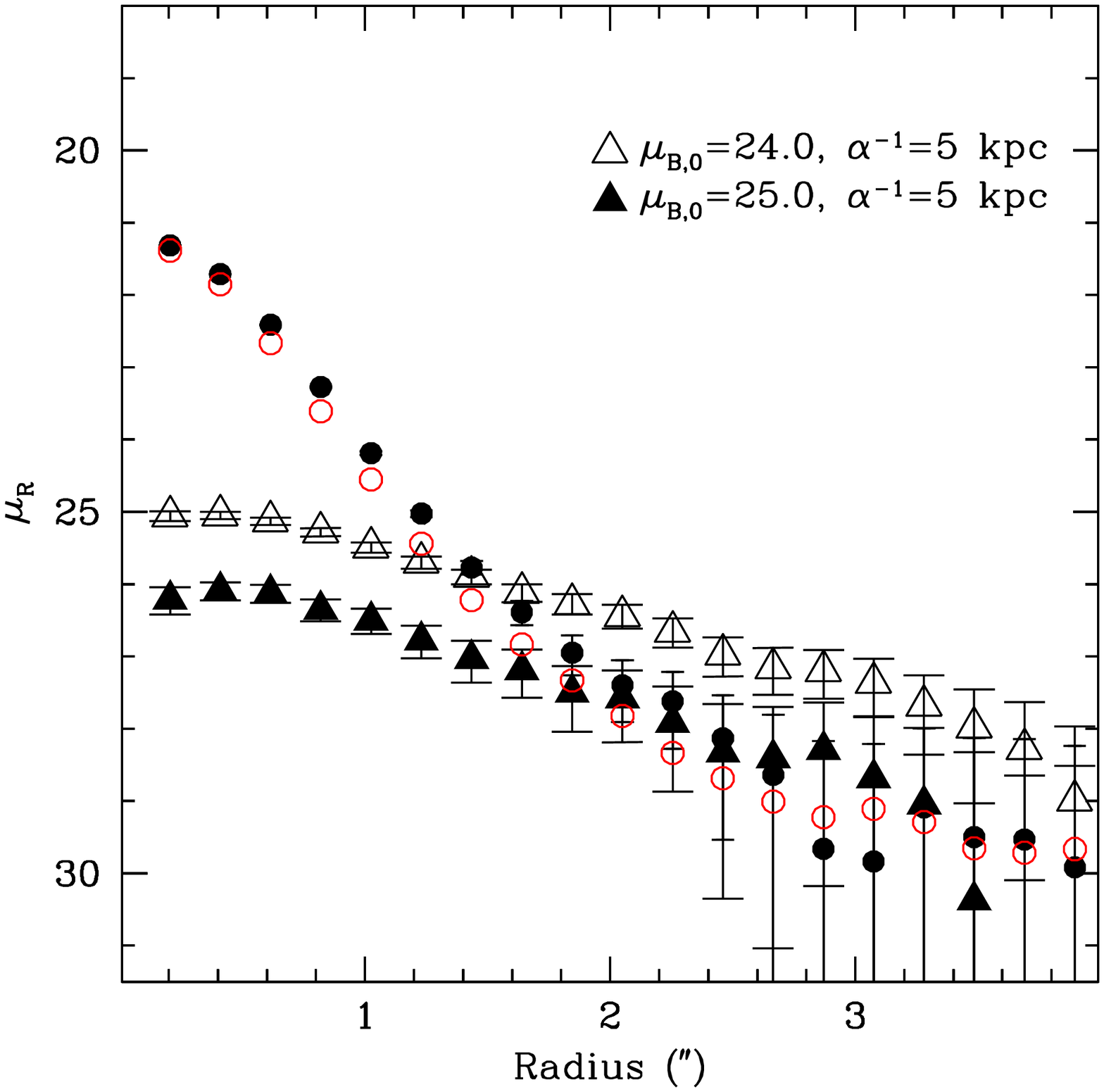}
\caption{Models of pure exponential disks underlying CNELGs. For
217169 at $z=0.354$, we plot circular surface brightness profiles of
the CNELG ({\it filled circles}) with nearby psf stars ({\it open red
circles}) and face-on disk models ({\it triangles, filled and open})
added to the same data.  Labels in the upper right indicate the
rest-frame central surface brightnesses and scale lengths of the disk
models.  We show models for two different disk sizes: a 2~kpc disk
({\it left}) and a 5~kpc disk ({\it right}).}
\label{fig:disks_tiny}
\end{figure}

The amount of possible fading we compute from the measured colors and
redshifts is not strongly correlated with our classification of the
surface brightness profiles of the CNELGs.  However, in \S 2.1 we
estimate that 1--3 of the CNELGs in our study are actually superposed
on flux from background or foreground galaxies.  Because these low
surface brightness features would not be evident in the spectra of
CNELGs, we cannot rule out the possibility that some of the apparently
larger CNELGs are actually superpositions.  However, the superposed
galaxies are not likely to affect the measured colors and the fading
analysis.  The two luminous galaxies that are roughly as compact as
tiny NGC 205 but that cannot fade to even M$_{\rm B} = -17$ are
surprising.  However, both galaxies are at relatively high redshifts,
where our size analysis has less discriminatory power; as a result,
both are also formally consistent with intermediate-sized dwarf
galaxies.

From this analysis, we conclude that most CNELGs (96 \%) could
conceivably fade to become at least as faint as luminous dwarf
galaxies ${\rm M_B > -17.4}$, although it is by no means guaranteed
that they will fade to these limits.  We predict that at most 11\% of
CNELGs fade to become very low-luminosity dwarf galaxies (${\rm
M_B} > -15.2$).\footnote{Accounting for the estimated errors in $B-V$,
this analysis shows that less than 44\% can fade to this limit given
the bluest possible colors consistent with the data.}  The remaining
CNELGs fade by smaller amounts to the present epoch, to become
sub-L$^{\star}$ galaxies or luminous dwarfs.  Thus, the
``bursting-faint-dwarf'' model works for {\it at most} 11 - 44\% of
CNELGs.  The others must fade to become higher-luminosity dwarfs or
more luminous galaxies.  In Sec.~3.2~--~3.4 we explore alternative
scenarios which involve more massive galaxy descendants.

\subsection{The bulges-in-formation hypothesis}

The analysis of \S~3.1 demonstrates that the bursting dwarf hypothesis
is not a satisfactory explanation for all of the CNELGs because some
CNELGs are too large to fade to dE's and some are not likely to fade
enough.  In \S~3.2~--~ 3.4, we explore the alternative possibility
that some CNELGs are actually starbursts in the centers of
intrinsically more massive galaxies that fade by $\lesssim 2$
magnitudes to become present-day luminous galaxies
\citep{Koo95,Kob99}.

Here, we explore the possibility that CNELGs are bursts of star
formation in the centers of spiral galaxies.  If these starbursts form
a substantial population of new stars in the centers of the galaxies,
they are almost certainly associated with the {\it in situ} formation
or enhancement of a galactic bulge.  In this scenario, the narrow
kinematic emission linewidths of the CNELGs may result from the fact
that the starburst is centrally concentrated.  The emission line flux
samples only the central part of the galaxy's potential well
\citep{Kob00, Bar01, Pis01}.  Similarly, CNELGs may be central
starbursts in disks viewed face-on \citep{Hom99,Hom02}.

Bulges in formation may be very difficult to discern.  The low
mass-to-light ratio of even a small forming bulge allows it to
dominate the flux of a spiral galaxy during rapid formation
\citep[e.g., Fig.~3 of][]{Bar01}.  In this scenario, the galaxy as
whole fades less to the present epoch, but it's surface brightness
distribution changes dramatically.  When the forming bulge fades, the
disk becomes much more obvious and the object appears as a disk galaxy
instead of a CNELG.

Because CNELGs are selected as compact objects, they are not embedded
in extremely obvious galactic disks.  With the exception of 212668,
which has an evident tidal tail in the new data, our deep ground-based
data do not reveal incontrovertible evidence for disks based on visual
inspection alone.  However, lower surface brightness disks and small
disks would not be evident in a visual inspection.  The analysis in
\S~3.1 demonstrates that only 4 of the CNELGs, those in the {\sc
large} category (that includes 212668), are more extended spatially
than even large dwarf galaxies.

As Fig.~\ref{fig:ngc205} illustrates using the \citet{Jong96} local
sample of spirals, most isolated disks at redshifts $\lesssim 1$ would
be resolved by our data.  However, the inner disks may be overwhelmed
by light from the central starburst in some CNELGs.  To be observed
clearly in a distant galaxy, the disk must exceed the surface
brightness limit of the observation at a radius well beyond the
influence of the central burst.  Thus, our ability to detect a disk
depends on both the central surface brightness of the disk {\it and}
its scale length: larger, higher surface brightness disks are easier
to see.  The ability to discern a disk is also a strong function of
redshift because of cosmological surface brightness dimming; distant
disks can fade into the sky noise without extending very far past the
influence of the bulge (or central starburst).  Finally,
disk detection depends on the luminosity of the central starburst ---
or equivalently the bulge-to-disk ratio --- because the luminosity of
the central source affects the radius at which its light dominates the
disk.  For these reasons, we explore the detectability of disks in our
CNELGs one at a time using models constructed at the redshift and
resolution appropriate for each CNELG.

\begin{figure}
\plottwo{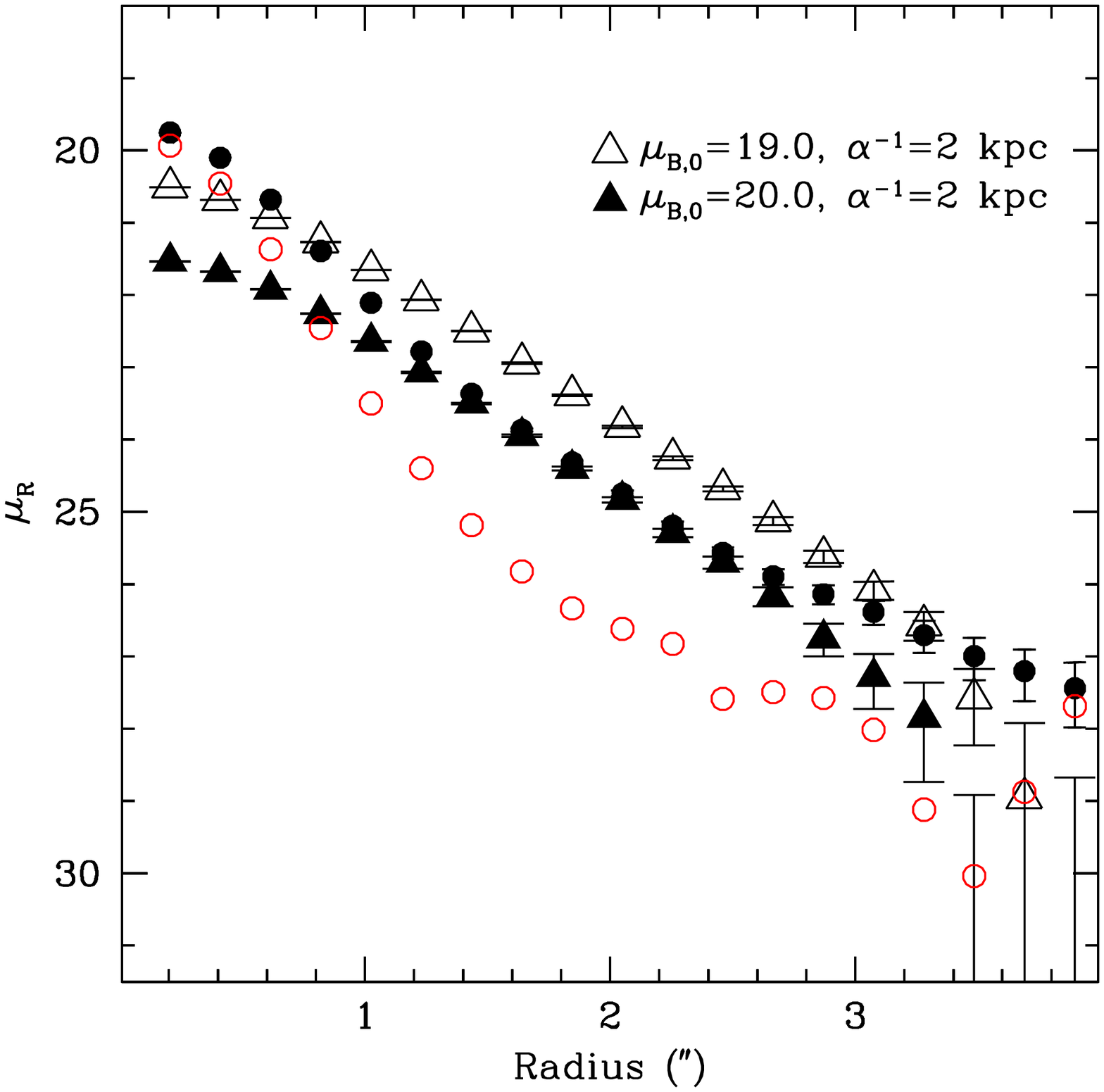}{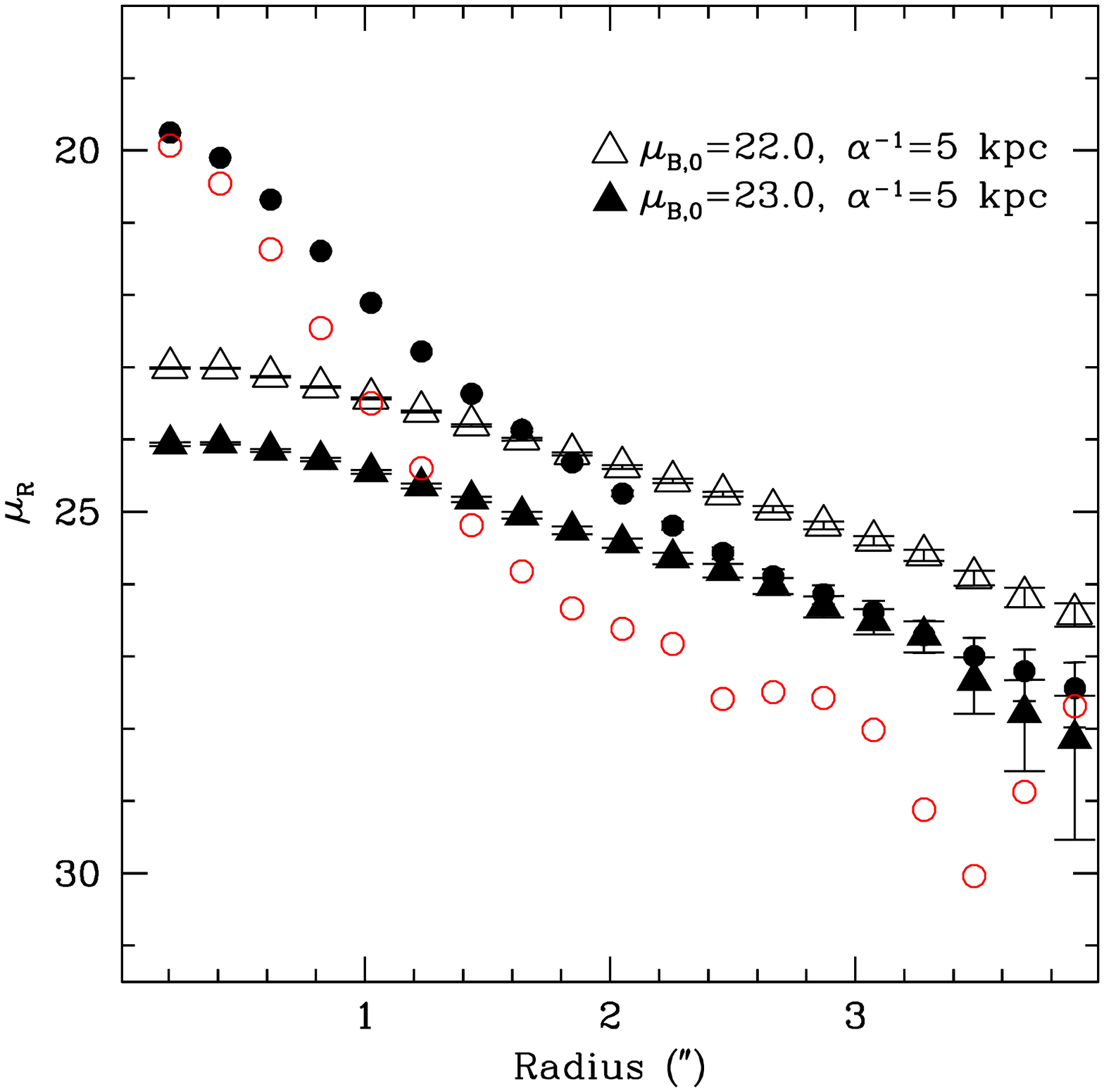}
\caption{Models of pure exponential disks, with symbols as in
Fig.~\ref{fig:disks_tiny}, but for different model parameters and for
the {\sc large} galaxy 217255 at $z=0.348$.}
\label{fig:disks_large}
\end{figure}

In the absence of a ``smoking gun'' for disks, we use our data to
explore the {\it possibility} that CNELGs are embedded in disks by
computing the surface brightnesses of the brightest disks of a given
size that could be accommodated by our data.
Figs.~\ref{fig:disks_tiny} -- \ref{fig:disks_large} show the circular
surface brightness profiles of two CNELGs plotted on top of circular
surface brightness profiles of models of isolated face-on disks with
Sbc-type spectral energy distributions that were added to blank-sky
regions of the data.  The plots are similar for moderately inclined
disks.  Highly inclined disks are likely ruled out because they would
be evident in the data, giving the galaxies a ``pinched'' appearance.

The figures show the sample disks near the maximal limits for a
marginally unresolved CNELG (Fig.~\ref{fig:disks_tiny}) and a
well-resolved, spatially extended CNELG (Fig.~\ref{fig:disks_large}).
Our data rule out disks that fall above the surface brightness of the
CNELG at large radii.  The results illustrated by
Fig.~\ref{fig:disks_tiny} are typical of the NGC 205-like CNELGs in
our study (see the classifications in Table~\ref{tab:CNELGs}).  These
CNELGs are not spatially extended enough to accommodate most of the
local disks larger than $\sim$2 kpc in the \citet{Jong96} sample.  The
detailed limits set by this analysis are a strong function of
redshift.  The lowest-redshift NGC 205-like CNELGs can accommodate
essentially no realistic face-on disks.  111255 requires $\mu_{\rm
B,0} \gtrsim 24$ for even a 2 kpc disk.

For example, in the left panel of Fig.~\ref{fig:disks_tiny}, we
illustrate that our data could accommodate a 2~kpc disk with $\mu_{\rm
B,0} = 22$ (for an Sbc-type spectral energy distribution), but not one
with $\mu_{\rm B,0} = 21$.  The right panel shows that the data
accommodate a 5~kpc disk with $\mu_{\rm B,0} = 25$, but not one with
$\mu_{\rm B,0} = 24$.  Substantially larger disks (e.g., 10 kpc) with
realistic surface brightnesses ($\mu_{\rm B,0} < 24$) are nearly
always ruled out.  Even Malin 1, with $\mu_{\rm B,0} \sim 25.5$ but
${\rm R_d = 78}$~kpc \citep{Im89}, would be quite evident in the data.

As Fig.~\ref{fig:ngc205} illustrates, the known disks typical in the
local universe occupy a limited range of surface brightnesses,
although this range may be artificially truncated by selection effects
even in the local universe.  While there are 2 kpc disks with
$\mu_{\rm B,0} > 21$, there are very few 5 kpc disks with $\mu_{\rm
B,0} \gtrsim 24$.  Thus, we conclude from Fig.~\ref{fig:disks_tiny}
that if the nearly unresolved CNELG 217169 is embedded in a disk, it
is either much lower surface brightness than most known disks in the
local universe or it is quite small ($\sim$2 kpc).

We compute the approximate maximum disk surface brightness accommodated
by the CNELGs.  We compare model disks (with noise) to the CNELGs in
the outer radii beginning where the model disk flux first exceeds the
CNELG flux.  We create models at 1-magnitude intervals in central
surface brightness and interpolate to the disk surface brightness that
corresponds to $\chi^2/{\rm (d.o.f.)} = 1$ in these outskirts.  We
list the results for Sbc-type disks in Table~\ref{tab:maxdisks},
sorted by redshift.  For two cases, we quote only a lower limit to the
maximum surface brightness (i.e., if 112259 has a 10~kpc disk, its
central surface brightness is $\mu_{\rm B,0} \gtrsim 27$). Assuming the
disks have bluer (Im-type) spectral energy distributions changes the
$B$-band surface brightnesses by $\leq 0.3$ magnitudes.  For context, 
we know that a the fiducial ``Freeman'' (bright) disk has a central 
surface brightness of $\mu_{\rm B} = 21.65$ \citep{freeman70}.

Both 111867 and 208846 accommodate 2 kpc disks with Sbc spectral
energy distributions only if $\mu_{\rm B,0} \gtrsim 23$; 111255
requires $\mu_{\rm B,0} \gtrsim 24.1$ for a 2 kpc and $\mu_{\rm B,0}
\gtrsim 26.2$ for a 5 kpc disk.  112259 requires $\mu_{\rm B,0} \geq
22.8$ for a 2 kpc disk and $\geq 26$ for a 5 kpc disk, and 217169
(Fig.~\ref{fig:disks_tiny}) requires $\mu_{\rm B,0} \gtrsim 22$ for a
2 kpc disk and $\mu_{\rm B,0} \gtrsim 24.3$ for a 5 kpc disk.  These 5
objects are the most unusual in nature.  All have $z < 0.4$ and
M$_{\rm B} \gtrsim -20$.  We discuss their nature further below.  The
higher-redshift NGC 205-like CNELGs accommodate more common small or
low surface brightness disks, as Table~\ref{tab:maxdisks} indicates.

\subsubsection{Implications of the maximum allowable disks}

The bulges-in-formation picture arose as a solution to the problem
suggested by the apparently high metallicities of the CNELGs that have
measured abundances.  \citet{Kob99} study the gas-phase metallicities
of a set of two CNELGs, finding that they lie along the
luminosity-metallicity relation of field galaxies. Although we note
that there are many exceptions to the luminosity-metallicity relation
such as NGC 205, the naive interpretation of their results may
restrict the amount of fading possible to these galaxies: the two
CNELGs in the Kobulnicky \& Zaritsky study appear to be too metal-rich
to fade to the observed low-luminosity (M$_{\rm B} \gtrsim -16$) dEs
in the local universe.  They predict fading for CNELGs that is
$\lesssim 1-2$ magnitudes.  The authors propose that CNELGs are either
progenitors to the more luminous and metal-enriched dEs, or they are
bulges forming before their disks are evident.  However, these
detailed metallicities have only been measured for a small number of
CNELGs.  In addition, the starbursts in CNELGs may represent a
relatively small fraction of the total gas at a later stage in the
galaxy's history, arguing that the metallicity in the gas phase may be
higher than the metallicity of the galaxy as a whole.

Although the present deep imaging data do not prove that most CNELGs
are embedded in galactic disks, we explore how the presence of a disk
could affect the fading of the CNELGs to the present day.  Here, we
hypothesize that the CNELGs are starbursts in the middle of face-on
disks that have the approximate maximum disk luminosities allowed by
the surface brightness profiles of the CNELGs
(Table~\ref{tab:maxdisks}).  Assuming the objects harbor Sbc-type
disks with the maximum allowed central surface brightnesses, most of
the 27 CNELGs are ``bulge-dominated'' (or starburst-dominated)
objects, with ${\rm \left<B/T\right> = 0.6}$ for the 2~kpc disks and
${\rm \left<B/T\right> = 0.8}$ for the 5~kpc disks.  For these maximum
allowable disks, these numbers are {\it lower limits} to the average
B/T (or starburst/T) ratios of the CNELGs.

If the disks do not evolve in luminosity or size but the central
starbursts fade substantially, the ${\rm B/T}$ ratios of the objects
decline rapidly \citep[see, e.g.,][]{Bar01}.  At the same time, the
total luminosities of the galaxies evolve, but by much less than the
fading central starbursts evolve.  In this way, the luminosity in the
disk ``stabilizes'' the CNELG against fading.  Here, we assume SEDs
for the disks and that the disks do not evolve substantially in
luminosity or color, then explore the fading caused by fading of the
``bulge'' component in two ways, by considering (1) the amount of
fading allowed to bring the galaxy's B/T ratio to a given final value
and (2) the amount of fading that will occur by the present day 
assuming the color of the observed CNELG must match the
color of the composite ``bulge''$+$disk at the redshift of the CNELG.

\begin{figure}
\plottwo{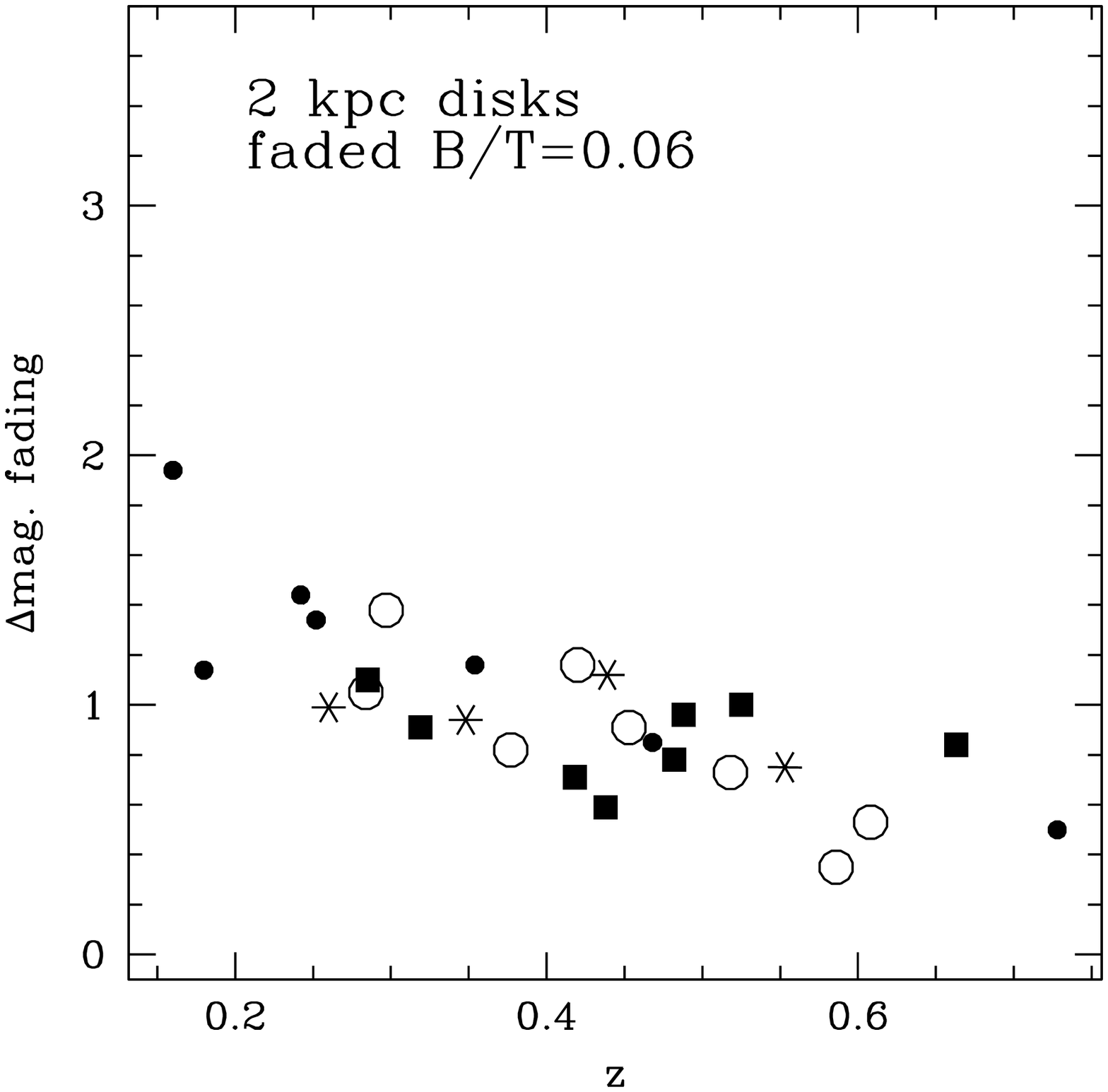}{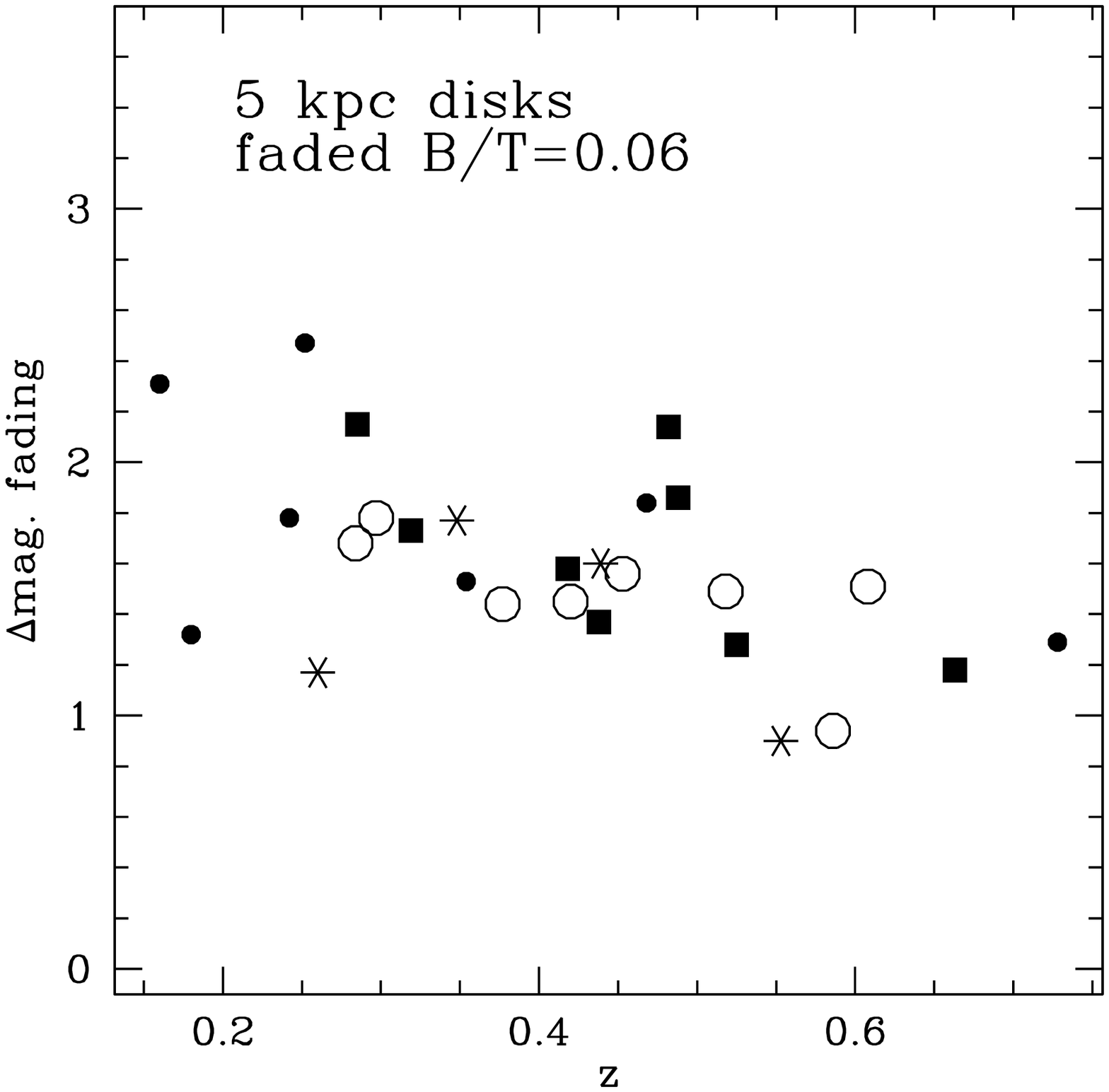}
\caption{Assuming the maximum allowable (face-on) disk described in
\S~3.2 and a non-evolving disk luminosity, we plot the amount of total
object fading required to achieve a ${\rm B/T = 0.06}$ for a 2 kpc
disk ({\it left}) or a 5 kpc disk ({\it right}) when the bulge is done
fading.  We show NGC 205-like ({\it filled circles}), {\sc
intermed}iate-class ({\it open circles}), VCC~437-like ({\it filled
squares}), and {\sc large} (disk-like) CNELGs ({\it asterisks}).}
\label{fig:2kpc}
\end{figure}

{\bf Fading to a given B/T ratio:} In Fig.~\ref{fig:2kpc}, we explore
the amount of total object fading that results from the fading of the
bulge to an eventual bulge-to-disk ratio that is more typical of
spiral galaxies.  We assume the maximum compatible disks described
above.  Thus, we show an approximate lower limit to the amount of
total object fading required to achieve a ${\rm B/T = 0.06}$ for the
faded object, which is appropriate for the progenitor of a typical
small late-type spiral \citep[see, e.g.,][]{Jong96}. The objects would
require less fading to achieve larger $B/T$ ratios appropriate for an
earlier-type spiral.  We plot the required fading to ${\rm B/T =
0.06}$ as a function of redshift to illustrate the mild systematic
trends with redshift that result from the heterogeneity of the sample
and our decreasing ability to rule out more luminous disks at higher
redshifts.  The classes of CNELGs from Table~\ref{tab:CNELGs} are
differentiated by point styles.

The relationship between $B-V$ color and amount of fading defined by
the models in Fig.~\ref{fig:bursts} allows us to examine whether this
picture is consistent with the colors of CNELGs.  In some cases, the
objects are too blue to support this picture: even if the disk is blue
(Im-type) and remains stable in color and $B$-band luminosity, the
bulges must fade to $B/T < 0.06$ to explain the color of the composite
CNELG.  

\begin{figure}
\plotone{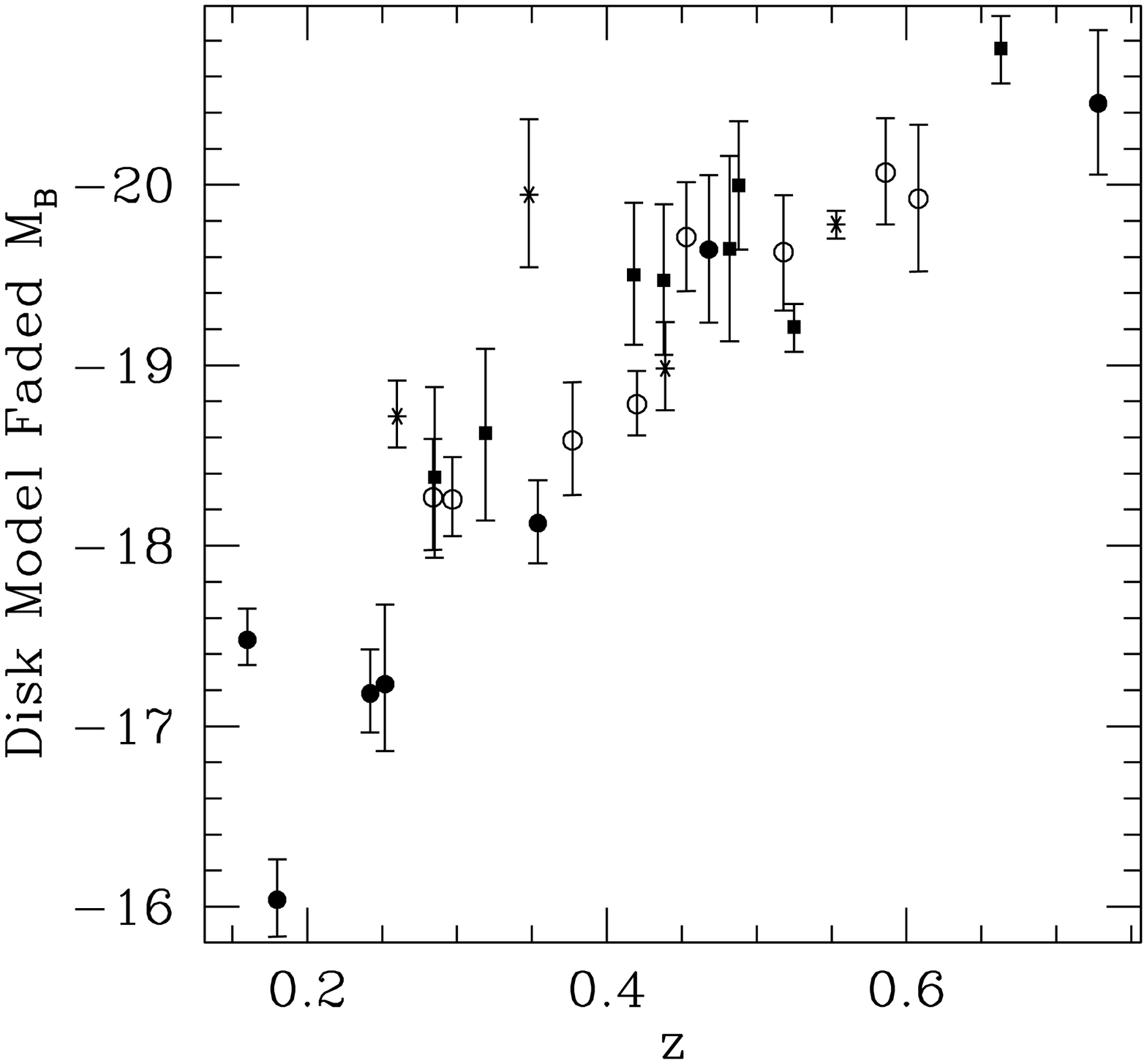}
\caption{The present-day luminosities of the CNELGs if they fade
according to the model in \S~3.2.1 based on maximizing the surface
brightness of potential ``hidden disks'' in which they are embedded.
The spread indicated by the error bars corresponds to the range of
results for different disk SEDs and for the 2~kpc disk and 5~kpc disk
models.   Point types are as in Fig~\ref{fig:colormag}.}
\label{fig:fade2}
\end{figure}

{\bf Color matching:} If we fix the color of the disk and match the
color of the composite object to the color of the actual CNELG,
allowing any final ratio $B/T \geq 0$, the objects themselves fade by
less than 2 magnitudes to match their colors if they are surrounded by
Sbc-type ($B-V=0.58$), Scd-type ($B-V=0.5$), or Im-type ($B-V=0.34$)
disks of maximal surface brightness.  We note that some of these disk
colors are not blue enough to allow a color equal to the object color
for a final, faded $B/T \geq 0$; in these cases we only allow the
bulge to fade to $B/T =0$.  Fig.~\ref{fig:fade2} shows the faded
magnitudes of the CNELGs in this model.  With the maximum allowed
disk, the objects fade little and most remain luminous; 89\% (24/27)
have mean faded M$_{\rm B} \leq -17.4$.  Conversely, the low-redshift
4 objects whose maximal fading is below M$_{\rm B} = -18$,
111867, 111255, 208846, and 112259, almost certainly cannot be luminous
galaxies.  Even without fading they have M$_{\rm B} > -19$.  These are
among the 5 objects identified in \S~3.1 as the best candidates for
starbursting dwarf galaxies.  The fifth, 217169, has a faded mean
M$_{\rm B} = -18.1$.  Because of the ``stabilizing'' disk population,
reddening can have little effect on Fig.~\ref{fig:fade2}.  Assuming
$B-V$ colors that are 0.3 magnitudes bluer changes the mean
results by an average of only 0.14 magnitudes for the 27 CNELGs.  If
the disks increase in luminosity through continued star formation, the
objects fade by even less than the predicted amount.  Conversely, if
the disks fade the objects will fade, but this effect is very general
and equally applicable to discussions of all disk galaxies at
intermediate redshift.

If most CNELGs actually harbor disks, they will fade by roughly
$\sim$1-2 magnitudes and most will remain luminous galaxies (M$_{\rm
B} \lesssim -18$).  As Fig.~\ref{fig:fade2} illustrates, the maximum
allowable small disks support enough stellar mass to stabilize most of
the galaxies against fading to dwarfs.  Only 4/27 CNELGs essentially
must fade to M$_{\rm B} \gtrsim -18$.

\subsection{The cores of elliptical galaxies in formation}

\begin{figure}
\plottwo{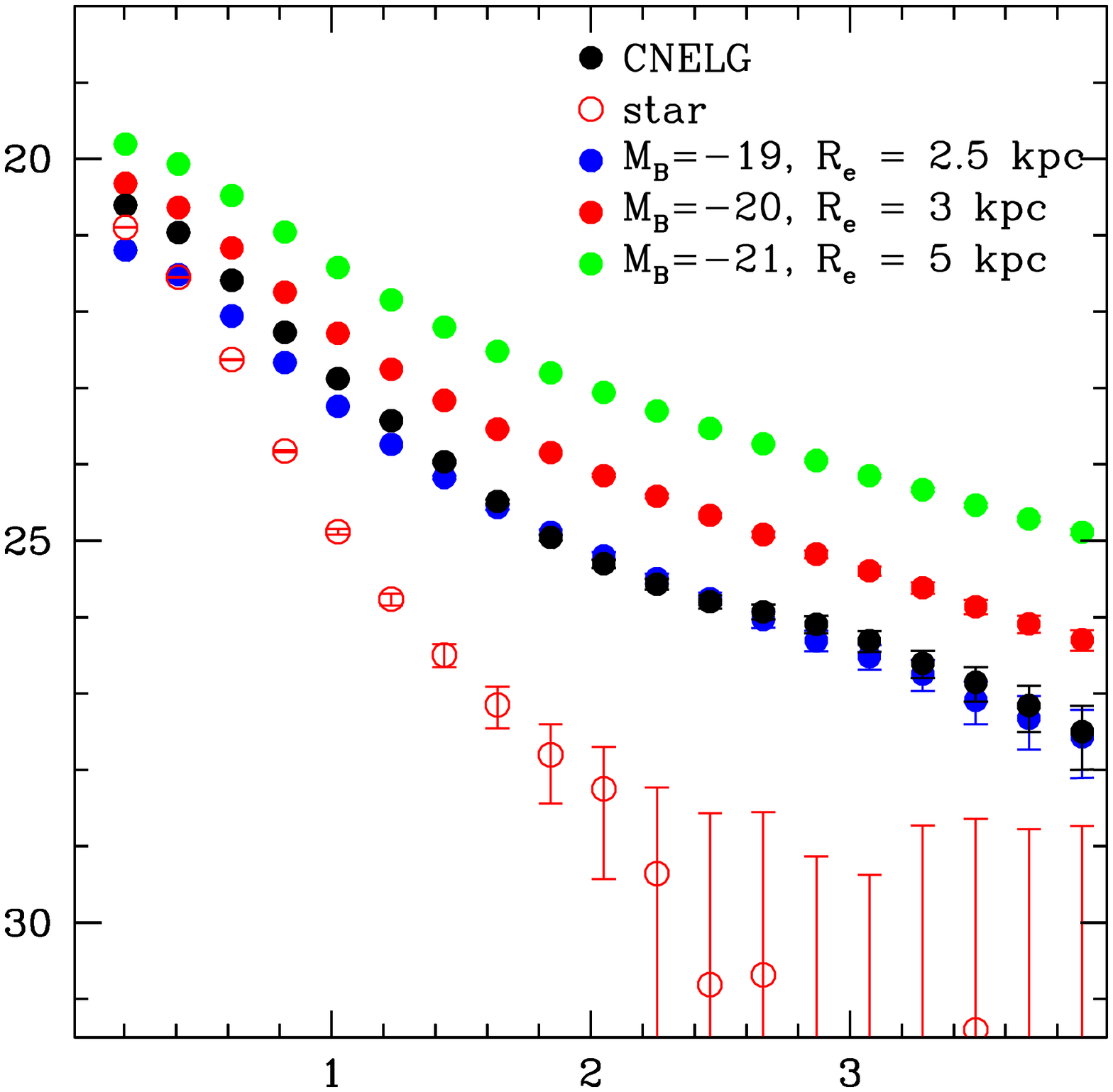}{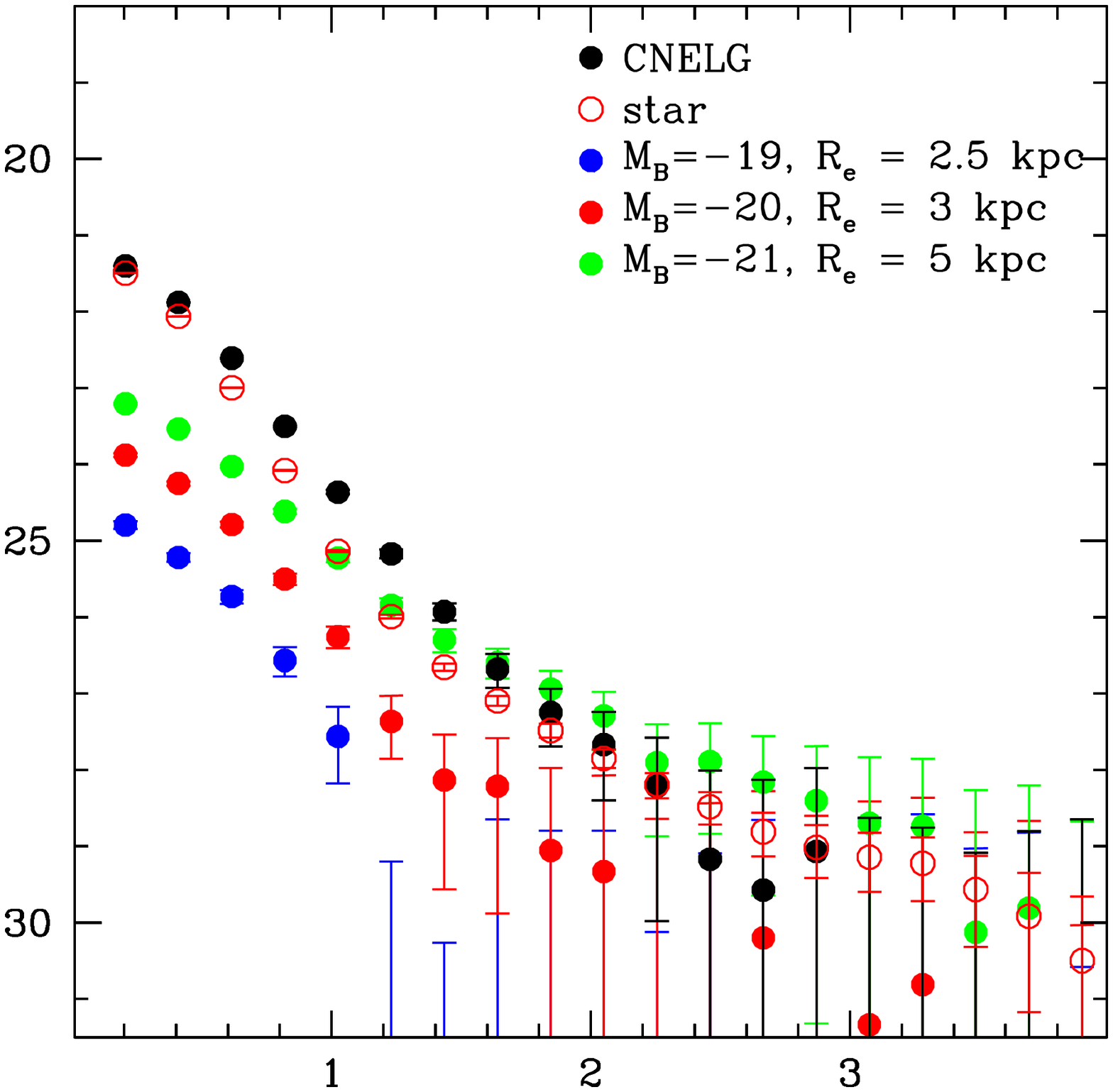}
\caption{Models of R$^{1/4}$-law elliptical galaxies underlying
CNELGs. For 212668 ({\it left}) at $z=0.260$ and 209640 at $z=0.728$
({\it right}), we plot circular surface brightness profiles of the
CNELGs ({\it filled black circles}) with nearby psf stars ({\it open
red circles}) and elliptical galaxy models ({\it colored points})
added to the same data.  Labels in the upper right indicate the
rest-frame luminosities and effective radii of the elliptical galaxy
models.}
\label{fig:ellip}
\end{figure}

The surface brightness profiles of most of the nearby CNELGs in our
study rule out the existence of a red underlying population with the
structural parameters of a giant elliptical.  However, the redshift
dependence in our ability to detect an underlying older galaxy is
extremely strong because of the much larger k-corrections associated
with older stellar populations.  In Fig.~\ref{fig:ellip}, we show the
surface brightness profiles of model elliptical galaxies on the
profiles for two CNELGs, 212668 ({\it left}), which is actually a {\sc
large} galaxy at lower redshift and 209640, a NGC 205-like galaxy at
the highest redshift of the sample, $z=0.728$.  We are unable to rule
out giant ellipticals from the surface brightness profile of the
distant galaxy, even though it is very compact.  However, very
luminous ellipticals are ruled out for the {\sc large} galaxy.  In
this example, our ability to find underlying red stellar populations
is dominated by redshift and not by the actual measured extent of the
galaxy.  Because of this extremely strong redshift dependence, the
surface brightness profile provides little  useful
information about the presence of a very old stellar population.

The strongest arguments against the hypothesis that CNELGs and other
compact blue galaxies at intermediate redshift are giant ellipticals
involve kinematics \citep[e.g.,][]{Im01}.  The velocity widths of
CNELG emission lines are narrower than the absorption line widths of
nearly all elliptical galaxies.  Concentrating the star formation in
the center of an elliptical to form a CNELG is unlikely to solve the
linewidth problem because many elliptical galaxies have velocity
dispersions that rise in the center \citep[e.g.,][]{Franx89, Im01}.
Some elliptical galaxies have a distinct central component organized
into a disk.  If the disk formed rapidly with a central burst of star
formation in the elliptical, the system might resemble a CNELG for a
brief time.  However, the nearby disks in ellipticals observed at high
spatial resolution appear to have similar star formation histories to
their parent galaxies \citep{Car97}.  Lower-luminosity elliptical
galaxies provide another possibility but the majority of even these
low-mass systems also have larger velocity dispersions than the CNELGs
\citep[e.g.,][]{Hal01}.

\subsection{CNELGs as progenitors of E+A galaxies}

By definition, E$+$A galaxies have strong Balmer absorption lines and
little or no line emission.  Thus, they are composed of a mix of older
stars and a dominant population of A-type stars, but no younger, more
massive stars; their spectra are consistent with a strong burst of
star formation that abruptly ceased $\sim$1 Gyr ago \citep{Couch87,
New90, Sch96, Gon99a, Gon99b}.  Although they are currently only a
small fraction of the luminous galaxy population
\citep[$\sim$0.2\%;][]{Zab96}, they represent a distinct phase of
post-strong-starburst evolution \citep{Dress83,Zab96,Quint04}.
Longslit spectroscopy demonstrates that the young components of the
stellar populations of E$+$A's are more centrally concentrated than
the older stars \citep{Nor01}.  Although detailed morphologies of the
separate young and old populations are not available, morphological
studies with HST of a subset of E$+$A's reveal disturbed galaxies
dominated by spheroids with small half-light radii; at least some of
the tidal features are low surface brightness and thus too difficult
to see at intermediate redshifts \citep{Yang04}. Thus, it is quite
possible that many, if not all, E$+$A galaxies are descendants of
CNELGs.  Because the definition of an E$+$A is independent of
morphology, this hypothesis and those outlined in \S~3.2 and \S~3.3
are not mutually exclusive.  To put it in another way, with a
sufficiently strong and short burst (e.g., akin to the bursting dwarf
scenario), CNELGs must go through an E+A phase.

High resolution images of a subset of 5 extremely blue non-cluster
E+A's show that they are spheroid dominated (${\rm B/T > 0.5}$),
although 2/5 E$+$A's with published HST images have spiral disks
\citep{Yang04}. Because kinematic studies of E$+$A's reveal little
rotation, \citet{Nor01} argue that E$+$A galaxies likely evolve into
dynamically hot, spheroid-dominated systems.  Thus, the CNELGs that
are progenitors to E$+$A's probably evolve into ellipticals or
early-type spirals.  Indeed, \citet{Ber05} find that 6/7
CNELGs with HST/STIS long-slit spectroscopy have V/$\sigma < 1$, i.e.,
in the spheroidal range.

Similar to the problems posed by identifying CNELGs with elliptical
galaxies (\S~3.3), the kinematics of the CNELGs continue to represent
a barrier to identifying them with E$+$A galaxies.  \citet{Nor01}
measure a range of velocity widths, $\sigma$, for the old stellar
populations of E$+$A's that place them only 0.6 magnitudes brighter
than the Faber-Jackson (1976) relation.  The measured values of
$\sigma$ vary from $\lesssim 23$ km s$^{-1}$ to 241 km s$^{-1}$.
Furthermore, the values of $\sigma$ measured from absorption lines
associated with the young stellar populations are on average {\it
larger} than those for the older populations.  The differences may
result from systematic differences between the kinematics of a young
emission-line population and its kinematics in absorption after
$\sim$1 Gyr of aging.  However, \citet{Kob00} find no evidence for
systematic differences between emission linewidths and absorption
linewidths in a sample of 22 local galaxies that span a wide range of
morphologies.  Alternatively, perhaps the only E$+$A's that go through
CNELGs phase are those with the narrowest linewidths, which is
possible if the E$+$A's with the narrowest linewidths are also the most
compact.  Existing studies of E$+$A's are not large enough to uncover
a trend of this sort.

\section{Discussion and Implications}

The two basic pictures for CNELG evolution can be categorized as (1)
the bursting dwarf hypothesis, where the galaxies are intrinsically
low-mass systems that fade to become present-day dwarfs, and (2) the
hypothesis that CNELGs are the centers of more massive galaxies that
fade little and remain luminous to the present day (${\rm L \gtrsim
0.1 L^{\star}}$).  \S~3.2-3.4 all discuss variants on the latter
hypothesis: that CNELGs are starbursts in the centers of disk
galaxies, possibly related to the formation of galactic bulges
(\S~3.2), that CNELGs are cores in elliptical galaxies (\S~3.3), and
that CNELGs are the immediate progenitors of E+A galaxies (\S~3.4).
Hypothesis (2) is very general, and may include many different types
of formation scenarios.

To explore the dwarf model, we calculate the {\it maximal} amount of
fading to the present day allowed by the lookback times and colors of
the CNELGs.  To explore the hypothesis that CNELGs are the centers of
more massive galaxies, we consider a scenario in which the CNELG is
surrounded by the largest exponential older (Im to Sbc-like) stellar
population accommodated by its surface brightness profile.  We assume
that the disk does not fade or change color with time and find the
amount of fading that matches the color.  Because the surrounding
older stellar population stabilizes the galaxy against fading, this
technique essentially allows us to compute a minimal amount of fading
in the case of a non-evolving disk.  If the disk increases in
luminosity with time, however, the object will fade less.

The surface brightness profiles we measure indicate that if CNELGs are
embedded in disks most of them are embedded in low surface brightness
disks.  We note, however, that this hypothesis is broadly consistent
with at least one favored formation scenario for the the bulge of the
Milky Way.  The bulge probably formed from a rapid burst of star
formation at early times \citep[e.g.,][]{Fer03}.  Although the disk
has a high surface brightness now, it has formed stars and increased
in surface brightness from much earlier times to the present epoch.
Thus, at the time the bulge formed, the disk may have been below the
central surface brightness limits allowed by Table~\ref{tab:maxdisks}.
Although the Milky Way bulge almost certainly formed long before
$z=0.728$, the basic formation scenario indicates that the Milky Way
may have gone through a CNELG-tye phase if its disk was low surface
brightness when its bulge formed.

\subsection{Sorting Different Types of CNELGs} 

Both the literature and our analyses support the conclusion that
CNELGs are a heterogeneous class of objects.  The CNELG sources
studied here were selected with three main attributes: (1) they are
unresolved sources in moderate-seeing ground-based images, (2) they
are selected from a magnitude limited sample; and (3) they are
selected to have blue colors unlike Galactic stars. The apparent
magnitude-limited nature of the sample results in a strong correlation
of luminosity with redshift. Roughly, sources above $z=0.4$ are
brighter than M$_B = -20$, while lower redshift sources are fainter
than M$_B = -20$. Surface-brightness limits on embedded disks (\S3.2)
are similarly correlated with redshift, with $\mu_{B,0,max}$ = 21 mag
arcsec$^{-2}$ corresponding to the break point at $z=0.4$. On the other
hand, because the color selection was narrowly defined and
k-corrections for these blue colors are small, there is little-to-no
trend in rest-frame color with redshift.  As Fig.~\ref{fig:fakedwarfs}
illustrates, even at the highest redshift in our sample we can
distinguish a small dwarf like NGC 205 from a large dwarf like VCC
437. We can also distinguish both types from much larger galaxies.
Most surprisingly, however, the size classes based on the analysis in
this paper are also not strongly correlated with redshift, with the
exception of the extreme compactness of the four lowest-redshift
CNELGs.  With these considerations of homogeneity and heterogeneity in
mind, we can combine constraints gleaned from the CFHT imaging
presented here with other information to gain insight into their
nature.

\begin{figure}
\plottwo{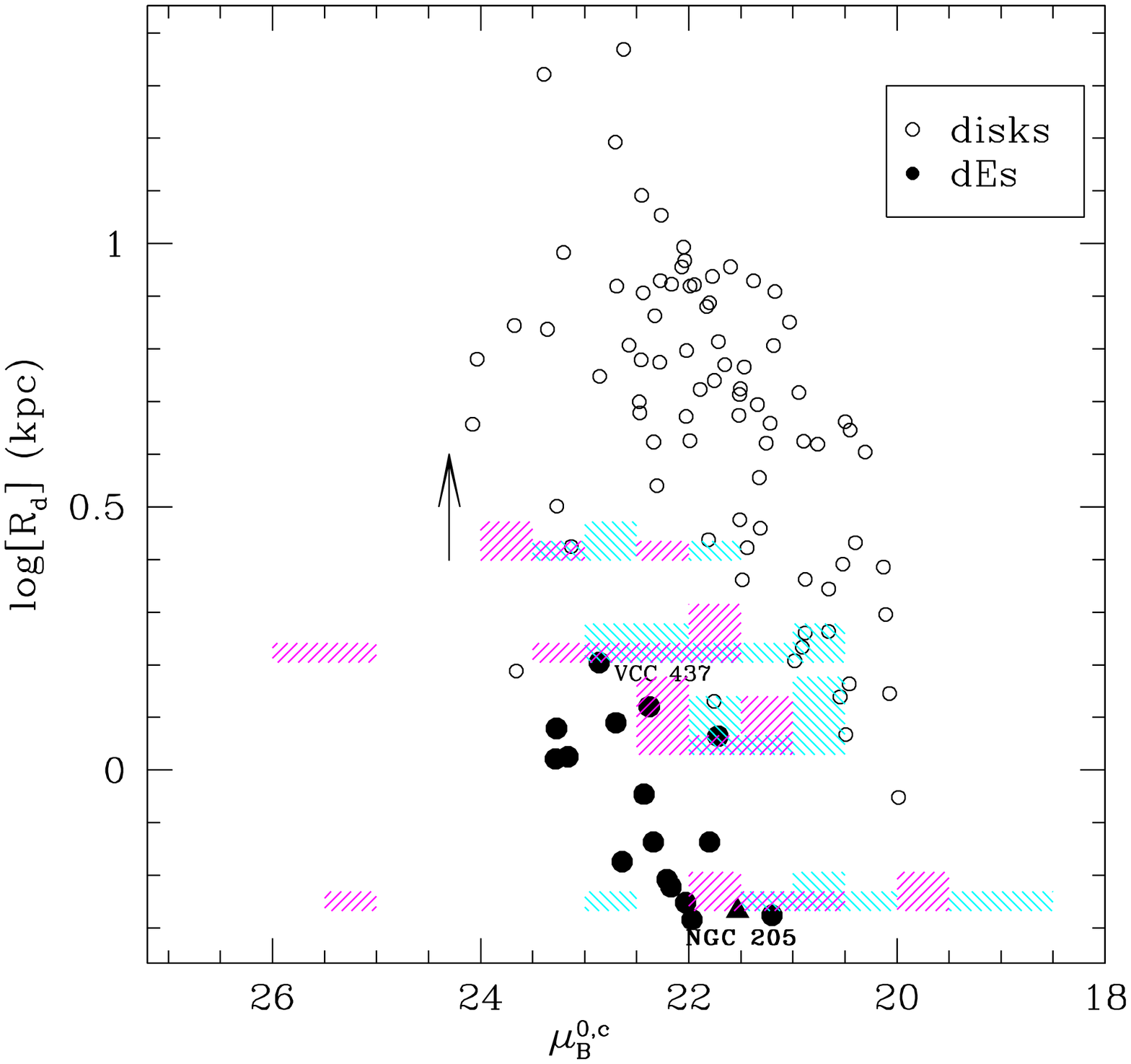}{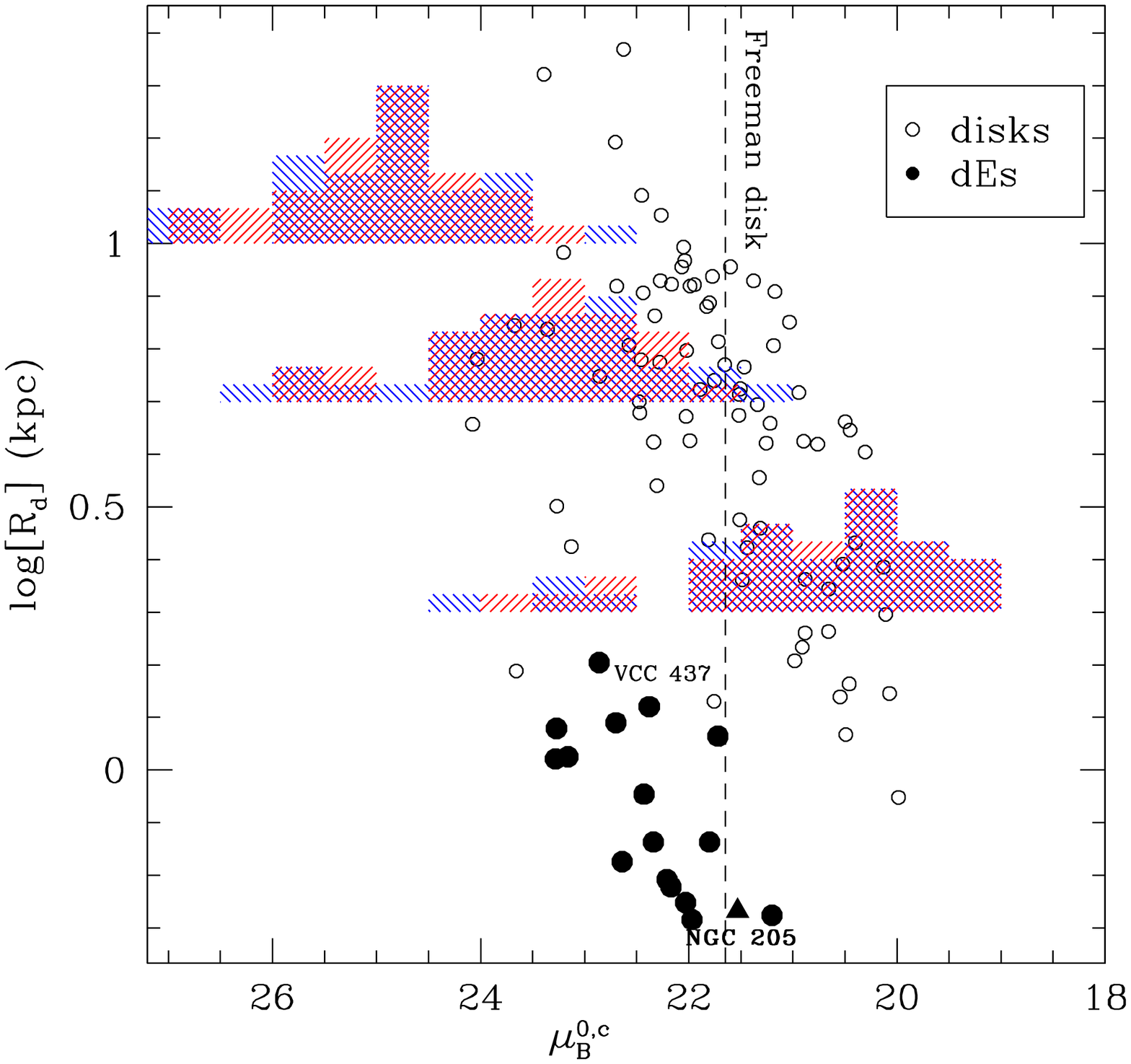}
\caption{Parameters of faded CNELGs in the maximal-fading model of
Fig.~\ref{fig:fade} ({\it left}; \S~3.1) and the minimal-fading model
corresponding to Fig.~\ref{fig:fade2} ({\it right}; \S~3.2) shown with
parameters for local dwarfs and disks as in Fig.~\ref{fig:ngc205}. For
the maximal-fading model, we assume exponential light profiles and
plot histograms of the dwarfs at their resulting central $B$-band
surface brightnesses of the faded dwarfs assuming the NGC 205-like,
INTERMEDiate, VCC 437-like, and LARGE dwarfs have sizes of 0.54, 1.07,
1.6, and 2.5 kpc, respectively.  The arrow indicates that the assumed
size of 2.5 kpc is an approximate lower limit for the LARGE class. The
magenta histograms are for fading including the reddening correction
and the blue assume no reddening.  For the minimal-fading model, we
plot the most luminous disks accommodated by the outer CNELG surface
brightness profiles at 2~kpc, 5~kpc, and 10~kpc (as in
Table~\ref{tab:maxdisks}).  The red and blue histograms assume Sbc and
Im spectral energy distributions, respectively. }
\label{fig:summary}
\end{figure}

We summarize the two basic models of dwarf galaxies from \S~3.1 and
\S~3.2 in Fig.~\ref{fig:summary}.  On the left, the colored histograms
correspond to the dwarfs of different sizes for the maximal fading
model. On the right, the colored histograms correspond to the
``maximal'' disks of 2,5, and 10 kpc accommodated by the surface
brightness profiles of the CNELGs (Table~\ref{tab:maxdisks}).

For the ``starbursting dwarfs'' model that assumes CNELGs will fade
maximally, Fig.~\ref{fig:summary}a shows histograms of the LARGE, VCC
437-like, INTERMEDiate, and NGC 205-like classes of CNELGs (top to
bottom).  All but the LARGE class of dwarfs overlap dEs in the local
universe.  However, for many of the most luminous CNELGs, the required 
{\it faded} central $\mu_B$ is brighter than typical dwarfs in the 
local universe, especially if we assume no reddening correction ({\it cyan
} histogram).  These objects either (1) are all reddened, (2) do not
fade as much as the maximal model suggests, or (3) fade to small
galaxies that are more luminous than the dwarf population that is typically
studied today.  The LARGE class of galaxies overlaps only with
the lower surface brightness small disks in this maximally-fading model.

For the ``hidden disks'' model that assumes CNELGs will fade
minimally, stabilized by an older population in a surrounding disk,
Fig.~\ref{fig:summary}b shows histograms of the brightest 2, 5, and
10 kpc disks accommodated by the data.  All but 4 CNELGs can harbor 2
kpc disks with parameters that overlap disks in the nearby universe.
In contrast, most cannot harbor 5 kpc disks with typical parameters,
and almost none can harbor 10 kpc disks.  Thus, if CNELGs are
starbursting centers embedded in typical galactic disks, their disks
are small.

The four lowest-redshift objects, 111867, 111255, 208846, and 112259,
are classified as NGC 205-like, suggesting that at redshifts below
$\sim$0.3, the extremely compact size requirements for the original
definition of CNELGs prohibit the selection of extended galaxies.  As
observed, they are already relatively low-luminosity, with $-18.8 \leq
{\rm M_B} \leq -17.2$.  Their colors, luminosities, and redshifts
suggest that within the errors they can fade to even M$_{\rm B}
\gtrsim -15$ (\S~3.2), and their surface brightness profiles and
colors suggest that they must fade to at least M$_{\rm B} \gtrsim
-17.5.$ Thus, the imaging data and the two models point to the same
interpretation: these 4 galaxies are starbursting dwarfs at low
redshift, observed at lookback times of $\lesssim 3$ Gyr.

Beyond $z=0.255$, the classifications of CNELGs appear more mixed.
Fig.~\ref{fig:fade} and Fig.~\ref{fig:fade2} show the approximate
maximum and minimum present-day luminosities of the CNELGs.  To within
the errors of Fig.~\ref{fig:fade}b, all but one of the CNELGs can fade
to M$_{\rm B} > -17.4$ and 8/23 CNELGs at $z > 0.255$ can fade to
M$_{\rm B} > -16$; the other 15 cannot.  In contrast, according to the
disk model in \S~3.2 (Fig.~\ref{fig:fade2}), the (approximate, mean)
maximum final luminosities of all the galaxies with $z > 0.225$ are
brighter than M$_{\rm B} = -17.4$.  Thus, all the CNELGs with $z >
0.255$ could be more luminous than the Virgo dEs at the present epoch
(based on the minimal-fading model).  Eight of these galaxies could
also fade to M$_{\rm B} > -15.2$ in the maximal-fading model.  The
remaining 15 must be more luminous than M$_{\rm B} = -15.2$ at the
present epoch.  A comparison of the fading limits derived in this
study with those of \citet{Guz98} show broad consistency with their
results.

From a morphological perspective, the more extended ({\sc large})
CNELGs are the most clear cut examples of luminous galaxies on the
basis of their sizes, including 212668, which has an evident tidal
tail in the $R$-band image.  Their colors allow a large range of
possible fading, however, indicating that three of the four can fade
to M$_{\rm B} > -16$ and that all 4 must fade to the range $-19.9 <
{\rm M_B} < -18.7$.  This LARGE sub-sample consists of only 4
of the 27 (14\%) of the CNELGs.  Moreover, we found in \S 2.1 that, on
a statistical basis, 1-3 of the CNELGs in our study may be superposed
on diffuse background galaxies.

The overlap between this study and the \citet{Kob99} metallicity study
includes only object, 206134, in the SA68 field, with $z=0.285$ and
M$_{\rm B} = -20.5$, which we classify as VCC 437-like based on its
surface brightness profile.  Its metallicity, $12+\log{\rm O/H} =
8.83$, lies directly on the luminosity-metallicity relation for nearby
star-forming galaxies.  The gas-phase abundance is high relative to
the stellar metallicity of some of today's dSph galaxies.  However,
other low-mass dwarfs such as the nucleated dE NGC 205 do have
comparable metallicity, and indeed other spheroidals such as
Sagittarius and Fornax lie significantly above the
luminosity-metallicity relation.  The model in Fig.~\ref{fig:fade}
allows a ``maximum'' fading to a very faint M$_{\rm B} = -13.7$ within
the errors on the color, and the model in Fig.~\ref{fig:fade2} yields
an ``average minimum'' fading to M$_{\rm B} = -18.4$, where
star-forming galaxies with its metallicity are quite typical in the
local universe \citep{Tre04}.

If many of the CNELGs really are progenitors to non-dwarf galaxies,
the observations of the kinematics of CNELGs remain the most difficult
to explain. Even some of the galaxies that are classified as {\sc
large} compared to dwarf galaxies have very narrow emission
linewidths.  These galaxies {\it consistently} exhibit narrow
linewidths that are observed nearby only in dwarf galaxies or in some
very low-luminosity ellipticals and bulges.  However, several authors
demonstrate that emission-line kinematics can be misleading
\citep{Kob00, Bar01, Pis01}, and that optical emission linewidths are
systematically low relative to 21 cm H{\sc I} for galaxies with narrow
optical emission linewidths \citep{Pis01}.

As discussed above, the linewidths observed in CNELGs are usually too
narrow for even lower-luminosity elliptical galaxies, even if the
emission-line flux is centrally concentrated \citep{Im01}.  However,
if the emission-line flux originates from small central disk, the
situation may be somewhat different.  Small circumnuclear disks have
been observed in both spiral and elliptical galaxies \citep[][and
references therein]{Rubin97, Car97}.  They are often the sites of
rapid star formation and, in spiral galaxies, may be associated with
the formation of a pseudobulge \citep[e.g.,][]{Kor04}.  These disks
exhibit a range of kinematics, however \citep{Rubin97}.  A lack of
observed rotational shear in medium spectral-resolution longslit
HST/STIS spectroscopy for 6/7 CNELGS argues against this picture
\citep{Ber05} -- except in the unlikely event that the disks are truly
face-on or the minor axis is consistently aligned with the
slit. However, at medium spectral-resolutions of 7000-10000, even
nearly face-on disks show rotational shear, e.g., as noted for the
starburst NGC 7673 \citep{Hom99}, or normal spirals \citep{And03}.
Our understanding of how to interpret kinematical data will likely
increase with future spectroscopic observations using high-resolution
bi-dimensional spectroscopy.

\section{Conclusions}

The compact narrow emission line galaxies (CNELGs) are an extreme,
rapidly evolving population of luminous galaxies observed at
intermediate redshift \citep{Koo94}.  Their nature and ultimate role
in the evolution of galaxies remains heavily debated.  We present deep
ground-based $R$-band imaging in good seeing designed to detect low
surface brightness features that may indicate CNELGs are starbursts
embedded in more massive systems.  We combine the new data with
existing color and redshift information to draw our conclusions:

\begin{enumerate}

\item We use circular-aperture surface photometry
to set lower limits to their present-day spatial extents irrespective
of luminosity, finding that $\sim$26\% are consistent with tiny dwarfs
like NGC 205 (or smaller, in some cases), 30\% are consistent with
intermediate-sized dEs, 30\% are consistent with larger dwarfs or
smaller disk galaxies, and 15\% appear more spatially extended than
dwarf galaxies.

\item Given the limits of the data, some CNELGs may harbor extended
structures such as galactic disks, although only 212668, with a clear
tidal tail, shows definitive evidence for one.  Our data indicate that
if most CNELGs harbor disks the disks must be small ($\lesssim$ a few
kpc in many cases) or relatively low surface brightness.

\item Models indicate that 4/27 (15\%), the four lowest-redshift
CNELGs, cannot harbor enough ``hidden'' disk material to stabilize
them against fading to become dwarf galaxies (M$_{\rm B} \gtrsim
-18$).  All of these galaxies are nearby ($z < 0.255$) and as small as
NGC 205.  They are likely examples of bursting dwarfs.

\item From an analysis of the maximum fading possible to CNELGs based
on their colors, we find that 15/27 (56\%) are not blue enough to fade
to tiny dwarf galaxies (M$_{\rm B} \geq -15.2$), even accounting for
reddening and errors.  These galaxies remain intrinsically more
luminous to the present day.  8/15 of these objects are classified as
NGC 205-like or intermediate-sized dwarfs based on their surface
brightness profiles.

\item From the same analysis of the maximum fading possible to CNELGs
based on their colors, we find that 26/27 (96\%) are blue enough to
fade to some type of dwarf galaxy (M$_{\rm B} \geq -17.4$), accounting
for reddening and errors.  However, four of these galaxies are too
spatially extended to become dwarfs.

\item From an analysis of the maximally bright disks accommodated by
the measured surface brightness profiles of CNELGs, all but the four
CNELGs can harbor small ($\sim$2 kpc) disks consistent with small
disks observed in the local universe.  Most cannot harbor large disks,
however.  An analysis of the minimal fading allowed to CNELGs based on
the potential disks they may harbor indicates that most CNELGs (85\%)
may fade by only a small amount to remain luminous (M$_{\rm B} < -18$)
at the present epoch.

\end{enumerate}

\acknowledgments

We thank David Koo, Richard Kron, Steve Majewski, Jeff Munn, and John
Smetanka for providing us with redshifts, photometry, and astrometry
for the CNELGs with us in advance of publication.  We thank Romeel
Dav\'{e} and James Bullock for helpful suggestions and insight.  We
thank John-David Smith for his insight into the process of unraveling
galaxy morphologies and for assistance that significantly improved our
software and data analysis. We thank Elizabeth Adams and Ian Roederer
for assisting in the observations of NGC 205.  Support for EJB was
provided by the Center for Cosmology at the University of California,
Irvine and by NASA through Hubble Fellowship grant \#HST-HF-01135.01
awarded by the Space Telescope Science Institute, which is operated by
the Association of Universities for Research in Astronomy, Inc., for
NASA, under contract NAS 5-26555; EJB and LvZ also acknowledge the
Herzberg Institute of Astrophysics, which is where this work was
begun.  LvZ acknowledges partial support from Indiana University and
NSF grants AST 03-47929 and AST 05-06054.  MAB acknowledges NASA/HST
AR grant 9917 and NSF grant AST-0307417, and gratefully acknowledges
the hospitality he enjoyed during his stay at the University of
Toronto, during which time the work was completed.

{
\renewcommand{\arraystretch}{1.3}
\begin{deluxetable}{lrcccccccccl}
\singlespace
\tabletypesize{\large}
\tablewidth{0pt}
\tablecolumns{12}
\tablecaption{The CNELGs}
\tablehead{
\colhead{} &
\colhead{} &
\colhead{Coordinates} &
\colhead{} &
\colhead{$z$} &
\colhead{} &
\colhead{} &
\colhead{} &
\colhead{$-\Delta m_{\rm max}$} &
\colhead{$\sigma$} &
\colhead{Exposure} &
\colhead{} \\
\colhead{Field} &
\colhead{ID} &
\colhead{(2000)} &
\colhead{$z$} &
\colhead{Quality} &
\colhead{m$_{\rm R}$} &
\colhead{M$_{\rm B}$} &
\colhead{$(B-V)_0$} &
\colhead{(mag.)} &
\colhead{(km s$^{-1}$)} &
\colhead{(ksec)} &
\colhead{Notes}}
\startdata
SA57 & 111867 & 13 07 14.0  $+$29 25 28 &  0.160 &  Q1 & 19.85 & -18.8 & 0.35 & 1.7(2.5) &    & 25.8 & NGC205 \\
SA57 & 111255 & 13 07 28.3  $+$29 24 40 &  0.180 &  Q3 & 21.78 & -17.2 & 0.16 & 2.7(6.1) &    & 25.8 & NGC205 \\
SA68 & 208846 & 00 16 31.1  $+$15 52 20 &  0.242 &  Q1 & 21.14 & -18.5 & 0.32 & 2.3(3.0) &    & 21.0 & NGC205 \\
SA57 & 112259 & 13 07 15.0  $+$29 26 03 &  0.252 &  Q3 & 21.06 & -18.7 & 0.33 & 2.3(3.1) &    & 25.8 & NGC205 \\
SA68 & 212668 & 00 18 04.9  $+$15 58 17 &  0.260 &  Q1 & 20.15 & -19.6 & 0.38 & 2.2(2.9) &    & 21.0 & LARGE \\
SA68 & 217054 & 00 17 40.8  $+$16 06 56 &  0.284 &  Q1 & 20.65 & -19.3 & 0.41 & 2.1(2.9) &    & 21.0 & intermed \\
SA68 & 206134 & 00 18 11.3  $+$15 47 54 &  0.285 &  Q1 & 20.29 & -19.8 & 0.27 & 2.6(3.6) & 41\tablenotemark{a} & 21.0 & VCC437 \\ 
SA68 & 203307 & 00 16 55.0  $+$15 43 06 &  0.297 &  Q1 & 20.59 & -19.5 & 0.35 & 2.4(3.1) &    & 21.0 & intermed \\
SA57 & 108954 & 13 09 35.7  $+$29 21 16 &  0.319 &  Q1 & 20.47 & -19.9 & 0.17 & 3.2(6.2) &    & 25.8 & VCC437 \\
SA68 & 217255 & 00 17 22.9  $+$16 07 26 &  0.348 &  Q1 & 19.40 & -21.2 & 0.31 & 2.7(3.4) & 40\tablenotemark{a} & 21.0 & LARGE \\ 
SA68 & 217169 & 00 16 56.2  $+$16 07 11 &  0.354 &  Q1 & 21.28 & -19.3 & 0.32 & 2.6(3.4) &    & 21.0 & NGC205 \\
SA68 & 217779 & 00 16 49.1  $+$16 08 47 &  0.377 &  Q3 & 21.18 & -19.6 & 0.40 & 2.4(3.2) &    & 21.0 & intermed \\
SA68 & 212916 & 00 16 28.9  $+$15 58 46 &  0.418 &  Q1 & 20.54 & -20.5 & 0.41 & 2.5(3.2) &    & 21.0 & VCC437 \\
SA57 & 112536 & 13 09 06.8  $+$29 26 32 &  0.420 &  Q1 & 21.07 & -20.0 & 0.32 & 2.8(3.5) &    & 25.8 & intermed \\
SA57 & 110601 & 13 08 47.7  $+$29 23 41 &  0.438 &  Q1 & 20.70 & -20.5 & 0.18 & 3.5(6.4) & 43\tablenotemark{b} & 25.2 & VCC437 \\ 
SA57 & 108956 & 13 09 48.2  $+$29 21 15 &  0.439 &  Q2 & 20.91 & -20.3 & 0.26 & 3.1(4.5) &    & 25.8 & LARGE \\
SA57 & 105482 & 13 09 08.7  $+$29 15 57 &  0.453 &  Q1 & 20.43 & -20.8 & 0.36 & 2.7(3.5) & 60\tablenotemark{b} & 25.8 & intermed \\ 
SA68 & 206403 & 00 17 17.6  $+$15 48 20 &  0.468 &  Q1 & 20.55 & -20.8 & 0.40 & 2.6(3.4) &    & 21.0 & NGC205 \\
SA68 & 215428 & 00 18 06.2  $+$16 03 17 &  0.482 &  Q2 & 20.60 & -20.9 & 0.42 & 2.5(3.3) &    & 21.0 & VCC437 \\
SA68 & 216388 & 00 18 16.5  $+$16 05 11 &  0.488 &  Q1 & 20.31 & -21.2 & 0.43 & 2.5(3.2) &    & 21.0 & VCC437 \\
SA57 & 117671 & 13 07 31.7  $+$29 34 33 &  0.518 &  Q2 & 21.07 & -20.6 & 0.44 & 2.5(3.1) &    & 25.8 & intermed \\
SA57 & 107042 & 13 07 26.2  $+$29 18 24 &  0.525 &  Q1 & 21.41 & -20.3 & 0.36 & 2.8(3.6) &116\tablenotemark{b} & 25.2 & VCC437 \\ 
SA68 & 217418 & 00 17 45.3  $+$16 07 47 &  0.553 &  Q1 & 21.20 & -20.7 & 0.30 & 3.0(3.8) &    & 21.0 & LARGE \\
SA57 & 118170 & 13 07 38.8  $+$29 35 26 &  0.586 &  Q2 & 21.30 & -20.8 & 0.35 & 2.9(3.7) &    & 25.8 & intermed \\
SA57 & 108945 & 13 08 37.8  $+$29 21 15 &  0.608 &  Q2 & 21.31 & -20.9 & 0.39 & 2.8(3.6) &    & 25.8 & intermed \\
SA57 & 117731 & 13 08 38.2  $+$29 34 48 &  0.663 &  Q1 & 20.65 & -21.8 & 0.31 & 3.1(3.9) &103\tablenotemark{b} & 25.8 & VCC437 \\ 
SA68 & 209640 & 00 17 55.3  $+$15 53 30 &  0.728 &  Q1 & 21.48 & -21.4 & 0.39 & 2.9(3.7) &    & 21.0 & NGC205 \\
\hline
\multicolumn{12}{c}{Compact Galaxies}\\
\hline
SA68 & 205935 & 00 17 41.2  $+$15 47 36 &  0.304 &  Q1 & 20.45 & -19.8 & 0.27 & 2.7(3.6) &   & 21.0 & LARGE \\
SA57 & 103605 & 13 07 53.0  $+$29 12 53 &  0.392 &  Q1 & 21.23 & -19.6 & 0.34 & 2.7(3.4) & 37\tablenotemark{b} & 25.2 & LARGE \\ 
SA57 & 104259 & 13 08 03.4  $+$29 14 00 &  0.601 &  Q1 & 21.01 & -21.2 & 0.38 & 2.8(3.6) & 52\tablenotemark{b} & 25.8 & LARGE \\
\hline
\enddata
\label{tab:CNELGs}
\tablecomments{Basic properties of the final sample of CNELGs and
compact galaxies: (1) Field, (2) ID number from the original study,
(3) J2000 coordinates, (4) redshift, (5) quality of redshift
information: Q1: high-quality redshift,Q2: one-line identification,
Q3: low-quality redshift, (6) measured $R$-band apparent magnitude,
(7) computed rest-frame $B$-band absolute magnitude, (8) computed
rest-frame $B-V$ color, (9) maximum fading in magnitudes of the CNELG
to the present day corresponding to the $B-V$ color without (with)
reddening correction and lookback time (see \S~3.1.1), (10)
emission-line velocity dispersion from the literature, (11) imaging
exposure time in $R$ for this study, and (12) classification from
\S~3.1: NGC205 = like NGC 205, intermed = between NGC 205 and VCC 437,
VCC437 = like VCC 437, and LARGE = larger than VCC 437 (disk
galaxy-like).}
\tablenotetext{a}{Guzman et al. (1996)}
\tablenotetext{b}{Koo et al. (1995)}
\end{deluxetable} }

{
\begin{deluxetable}{lrrrr}
\tabletypesize{\large}
\doublespace
\tablecolumns{5}
\tablecaption{Maximal $\mu_{\rm B,0}$ for fixed-size disks}
\tablehead{
\colhead{} &
\colhead{} &
\colhead{\rm $\mu_{B,0,max}$} &
\colhead{\rm $\mu_{B,0,max}$} &
\colhead{\rm $\mu_{B,0,max}$} \\
\colhead{Field} &
\colhead{ID} &
\colhead{\rm 2 kpc disk} &
\colhead{\rm 5 kpc disk} &
\colhead{\rm 10 kpc disk}}
\startdata
SA57O & 111867 & 23.3 & 25.7 & 26.6 \\
SA57N & 111255 & 24.1 & 26.2 & $>$27.0 \\
SA68E & 208846 & 23.1 & 25.4 & 26.5 \\
SA57P & 112259 & 22.8 & 26.0 & $>$27.0 \\
SA68F & 212668 & 21.5 & 23.7 & 25.2 \\
SA68H & 217054 & 21.9 & 24.5 & 25.9 \\
SA68C & 206134 & 21.5 & 24.5 & 25.9 \\
SA68A & 203307 & 22.0 & 24.4 & 25.6 \\
SA57K & 108954 & 21.1 & 23.9 & 25.3\\
SA68J & 217255 & 19.9 & 22.7 & 24.2 \\
SA68I & 217169 & 22.0 & 24.3 & 25.6 \\
SA68L & 217779 & 21.4 & 24.0 & 25.7 \\
SA68G & 212916 & 20.4 & 23.2 & 24.8 \\
SA57Q & 112536 & 21.3 & 23.6 & 24.9 \\
SA57G & 110601 & 20.2 & 23.0 & 24.7 \\
SA57L & 108956 & 21.0 & 23.4 & 24.7 \\
SA57D & 105482 & 20.2 & 22.8 & 24.3 \\
SA68D & 206403 & 20.2 & 23.2 & 24.8 \\
SA68N & 215428 & 20.1 & 23.4 & 25.3 \\
SA68O & 216388 & 19.9 & 22.8 & 24.6 \\
SA57R & 117671 & 20.3 & 23.0 & 24.5 \\
SA57F & 107042 & 20.8 & 23.1 & 24.7 \\
SA68K & 217418 & 20.2 & 22.4 & 23.9 \\
SA57S & 118170 & 19.7 & 22.3 & 24.0 \\
SA57J & 108945 & 19.7 & 22.7 & 24.4 \\
SA57H & 117731 & 19.1 & 21.5 & 22.9 \\
SA68M & 209640 & 19.2 & 22.0 & 23.6 \\
\hline 
\multicolumn{5}{c}{Compact Galaxies} \\
\hline 
SA68B & 205935 & 21.2 & 23.8 & 25.3 \\
SA57B & 103605 & 20.7 & 23.6 & 25.0 \\
SA57C & 104259 & 19.3 & 22.0 & 23.6 \\
\hline
\enddata
\tablecomments{The maximal central surface brightness of disks
accomodated by our data:  (1) Field, (2) ID number, and 
maximal $B$-band surface brightness for a disk with 
an Sbc-type spectral energy distribution and a scale 
length of (3) 2 kpc, (4) 5 kpc, and (5) 10 kpc.  We note that
maximal central surface brightnesses become substantially
less accurate at $\mu \sim 27$ for 5 kpc disks and 
and at $\mu \sim 25$ for 10 kpc disks.}
\label{tab:maxdisks}
\end{deluxetable} }


\end{document}